\documentclass[a4paper,11pt]{article}

\usepackage{amsmath,amssymb,amsthm}
\usepackage{authblk}  
\usepackage[british]{babel}
\usepackage{bbm}  
\usepackage[sorting=none,doi=false,isbn=false,giveninits=true,style=nature,url=false]{biblatex}           
\usepackage{caption}
\usepackage{color}
\usepackage[utf8]{inputenc}  
\usepackage{csquotes}
\usepackage[T1]{fontenc}
\usepackage[top=2.5cm, bottom=2.5cm, left=2.54cm, right=2.54cm]{geometry}
\usepackage{graphicx}
\PassOptionsToPackage{hyphens}{url}\usepackage[colorlinks = true,
            linkcolor = magenta,
            urlcolor  = blue,
            citecolor = blue,
            anchorcolor = blue]{hyperref}
\usepackage{cleveref}
\usepackage[mono=false]{libertine}
\usepackage{nicefrac}
\usepackage{microtype}      
\usepackage{physics}
\usepackage{subcaption}
\usepackage{tabularx}
\usepackage{wrapfig}
\usepackage{xcolor}

\usepackage{amsmath,amsfonts,bm}

\newcommand{\figleft}{{\em (Left)}}
\newcommand{\figcenter}{{\em (Center)}}
\newcommand{\figright}{{\em (Right)}}
\newcommand{\figtop}{{\em (Top)}}

\def\1{\bm{1}}

\def\va{{\bm{a}}}
\def\vb{{\bm{b}}}

\def\vu{{\bm{u}}}

\def\vx{{\bm{x}}}

\def\vz{{\bm{z}}}

\def\eva{{a}}
\def\evb{{b}}

\def\evu{{u}}

\def\evx{{x}}

\def\mA{{\bm{A}}}

\def\mI{{\bm{I}}}
\def\mJ{{\bm{J}}}

\def\mM{{\bm{M}}}

\def\mO{{\bm{O}}}

\def\mS{{\bm{S}}}

\def\mW{{\bm{W}}}
\def\mX{{\bm{X}}}
\def\mY{{\bm{Y}}}

\DeclareMathAlphabet{\mathsfit}{\encodingdefault}{\sfdefault}{m}{sl}
\SetMathAlphabet{\mathsfit}{bold}{\encodingdefault}{\sfdefault}{bx}{n}

\def\emA{{A}}

\def\emJ{{J}}

\def\emM{{M}}

\def\emS{{S}}

\def\emW{{W}}
\def\emX{{X}}

\DeclareMathOperator\erfc{erfc}
\DeclareMathOperator{\mse}{\sf mse}

\newcommand{\out}{\text{out}}

\bibliography{reconstructing}

\definecolor{C0}{HTML}{1f77b4}
\definecolor{C1}{HTML}{ff7f0e}
\definecolor{C2}{HTML}{2ca02c}
\definecolor{C3}{HTML}{d62728}

\title{Bayesian reconstruction of memories\\ stored in neural networks from their
  connectivity}

\author[1]{Sebastian Goldt\thanks{sgoldt@sissa.it}}
\author[2]{Florent Krzakala}
\author[3]{Lenka Zdeborov\'a}
\author[4,5]{Nicolas Brunel}
\affil[1]{International School of Advanced Studies (SISSA), Trieste, Italy}
\affil[2]{IdePHICS laboratory, \'Ecole F\'ed\'erale Polytechnique de Lausanne (EPFL), Switzerland}
\affil[3]{SPOCS laboratory, \'Ecole F\'ed\'erale Polytechnique de Lausanne (EPFL), Switzerland}
\affil[4]{Department of Neurobiology, Duke University, Durham, NC, USA}
\affil[5]{Department of Physics, Duke University, Durham, NC, USA}

\date{\today}

\begin{document}

\maketitle

\begin{abstract}

The advent of comprehensive synaptic wiring diagrams of large neural circuits
has created the field of connectomics and given rise to a number of open
research questions. One such question is whether it is possible to reconstruct
the information stored in a recurrent network of neurons, given its synaptic
connectivity matrix.  Here, we address this question by determining when solving
such an inference problem is theoretically possible in specific attractor
network models and by providing a practical algorithm to do so. The algorithm
builds on ideas from statistical physics to perform approximate Bayesian
inference and is amenable to exact analysis. We study its performance on three
different models, compare the algorithm to standard algorithms such as PCA, and
explore the limitations of reconstructing stored patterns from synaptic
connectivity.

\end{abstract}

\section*{Introduction}
\addcontentsline{toc}{section}{Introduction}

Comprehensive synaptic wiring diagrams or ``connectomes'' provide a detailed map
of all the neurons and their interconnections in a brain region or even an
entire organism. Since the connectome of the nematode C.~elegans was obtained
using electron microscopy methods in 1986~\cite{White1986}, methods for data
acquisition and analysis have both been scaled up and improved
significantly. Today, it has become possible to provide connectomes of much more
complex systems such as various \emph{Drosophila melanogaster}
circuits~\cite{Ohyama2015,Eichler2017a}, or even a large part of its
brain~\cite{xu20,scheffer20}; the olfactory bulb of zebrafish~\cite{wanner20};
and various pieces of the rodent 
retina~\cite{briggman2011wiring, Helmstaedter2013, kim2014space},
hippocampus~\cite{Mishchenko2010}, and
cortex~\cite{kasthuri2015saturated, Lee2016a, Motta2019a, dorkenwald19,macrina21,bae21}. While
there still remain a number of formidable challenges on the way to the complete
connectome of a mammal or even human brain~\cite{Motta2019b}, the data sets
available today already give rise to a number of intriguing questions. At the
same time, it is becoming increasingly clear that new quantitative methods must
be developed to fully exploit the new troves of data that connectomics
provides~\cite{Litwin-Kumar2019}.

Here, we focus on local neural networks that store information in their synaptic
connectivity. It has been hypothesised that cortical networks are optimised for this
task, thanks to their
extensive recurrent synaptic connectivity ~\cite{Brunel2016}. A popular model for these networks are attractor neural
networks such as the Hopfield model~\cite{hopfield1982} and various
generalisations~\cite{tsodyks1988a, Amit1995, amit1997model, pereira2018attractor},
in which memories are stored as attractor states of the dynamics. These attractor
states represent learned internal representations of external stimuli that have
been presented repeatedly to the network during training, inducing changes in
synaptic weights of the network. One natural question to ask is then: given the
knowledge of the synaptic connections between neurons in a recurrent neural
network, can we reconstruct the patterns of activity that were stored in this
network in the first place?

In this paper, we first give a mathematical formulation of this problem in terms
of a Bayesian inference problem. The Bayesian approach has the advantage of
providing a natural way to handle the uncertainty associated with estimating a
large number of parameters from a large number of noisy observations,
\emph{i.e.} the reconstruction of the original stimuli from the strength of the
synapses in the networks in the present case. Modelling the noise is crucial in
this problem, as we cannot expect the synaptic strengths reported in connectomes
to be more than rough estimates. We use tools from statistical physics to
both determine when solving this inference problem is theoretically possible in
a model setting, and to provide a practical algorithm to do so. We analyse the
performance of the algorithm in detail on a variety of different problems, and
we invite the reader to download our reference implementation of the algorithm
on \href{https://github.com/sgoldt/reconstructing_memories}{GitHub} and to use
and extend it.

\subsection*{The task: Reconstructing memories from network connectivity}
\addcontentsline{toc}{subsection}{The task: Reconstructing memories from network connectivity}

\subsubsection*{The network model} 
\addcontentsline{toc}{subsubsection}{The network model}

We analyse a variant of the celebrated Hopfield
model~\cite{hopfield1982} for a recurrent neural network composed of $N$
interacting neurons with state $s_i$, $i=1,\ldots,N$. The network is
fully-connected with symmetric, bidirectional connections that have a scalar
weight $\emJ_{ij}=\emJ_{ji}\in\mathbb{R}$. The neurons update their state at
iteration $k+1$ sequentially according to 
\begin{equation}
    \label{eq:dynamics}
    s_i^{k+1} = g(a_i^k),
\end{equation}
where $g(\cdot)$ is some non-linear activation function and
$a_i^k = \sum_{j\neq i}\emJ_{ij}s_j^k$ is the total synaptic input of the $i$th
neuron.

The network stores $P$ fixed patterns or memories, which are
$N$-dimensional vectors that we collect in the matrix
$\mX^* \in \mathcal{X}^{P\times N}$. We write $\mX^*_{\mu, :}$ for the $\mu$th
pattern stored in the network, and $\mathcal{X}$ denotes the set of
values that pattern entries can take, \emph{e.g.}  $\mathcal{X}=\{\pm 1\}$ for
binary patterns. Note that these patterns correspond to deviations of neuronal activity from its mean, and not neuronal activity itself, which is constrained to be non-negative. This subtraction of mean activity is expected to be performed by the plasticity rule operating at the synapse (using a `covariance rule', see e.g.~\cite{dayan01}). In the case of binary patterns,
$X_{\mu,i}=1$ means a neuron is active in a given pattern, while $X_{\mu,i}=-1$ means a neuron is inactive.
We also assume that $X_{\mu,i}$ are i.i.d. random variables, i.e.~that stored patterns are uncorrelated.
Patterns are stored in the network by choosing its weights
$\emJ_{ij}$ such that the patterns $\mX^*_{\mu, :}$ become fixed points of its
dynamics.

We study this model in the thermodynamic limit $N\to\infty$, while keeping the
number of patterns $P$ of order 1. This scaling makes the resulting weight
matrix of the Hopfield model a low-rank matrix. Low rank matrices have played an important role in
neuroscience in recent years, in particular for the modelling of recurrent
networks~\cite{mastrogiuseppe2018linking, schuessler2020interplay,
  sussillo2009generating, logiaco2021thalamic}. Hence, methods to estimate them
from data in a principled way can help connect these theories with experimental
data.

\subsubsection*{The learning rule} 
\addcontentsline{toc}{subsubsection}{The learning rule}

A classic idea for choosing the weights, or the
connectivity structure of the network
$\mJ=[\emJ_{ij}]\in\mathbb{R}^{N\times N}$, is to choose the weights
proportional to the empirical correlation of the patterns,
$\emW_{ij} \sim \sum_\mu \emX^*_{\mu, i} \emX^*_{\mu, j}$. This prescription is
also known as the Hebb rule~\cite{hebb1949} and can be written more compactly as
\begin{equation}
  \label{eq:w}
  \mJ=\mW=\frac{1}{\sqrt{N}} \mX^* \left(\mX^*\right)^\top,
\end{equation}
where $\left(\mX^*\right)^\top$ is the transpose of $\mX^*$ and $\mW$ is thus
the empirical correlation matrix of the patterns, assuming that the means of the
patterns are zero, as we will do throughout this work, and the choice of the $1/\sqrt{N}$ scaling is explained below. With binary neurons
$s_i(t)=\pm 1$, this model corresponds to the celebrated Hopfield
model~\cite{hopfield1982}. In this model, the network exhibits fixed point
attractors close to the stored memories, provided the number of stored patterns
$P$ is smaller than $\alpha_c N$ where
$\alpha_c\sim 0.14$~\cite{amit1987statistical}. 

The connectivity matrix \cref{eq:w} has a number of unrealistic features that
makes it inadequate for the problem we are interested in here: (i) The network
is fully connected, at odds with neuronal networks in the brain; (ii) Synapses
are not sign-constrained, while synapses in the brain are either excitatory
(i.e.~non-negative) or inhibitory (i.e.~non-positive). A minimal model that
satisfies both requirements is the \textbf{rectified Hopfield model}
\begin{equation}
  \label{eq:J}
  \emJ_{ij} = \Phi (\emW_{ij} - \tau + \zeta_{ij}) \ge 0
\end{equation}
where $\tau>0$, $\zeta_{ij}$ is a noise term (see below), and we choose
$\Phi(x) = \max(0, x)$. This choice ensures that weights are non-negative,
effectively yielding a model of a network of excitatory neurons. This is
consistent with the hypothesis that information storage occurs primarily in
excitatory-to-excitatory synapses, while the job of inhibitory neurons is
primarily to control the level of activity in the excitatory network. This view
is consistent with a number of studies~\cite{grienberger17, lim2015inferring}
but has been challenged by others~\cite{mongillo2018inhibitory}. 

The noise term $\zeta_{ij}$ is taken to be a symmetric random Gaussian matrix,
i.e.~for $i<j$,~$\zeta_{ij}$s are i.i.d.\ random Gaussian variables with mean
zero and standard deviation~$\nu$, and~$\zeta_{ji}=\zeta_{ij}$.  The scalar
parameter $\tau$ controls the connection probability in the network,
\begin{equation}
  \label{eq:p_C}
  p_C = p(J_{ij} > 0) = \frac{1}{2}\erfc\left(\frac{\tau}{\sqrt{2}\nu}\right)
\end{equation}
to leading order as $N\to\infty$. In particular, the network becomes sparse in
the large $\tau$ limit. Since $P\sim O(1)$, the weights obtained from
\cref{eq:w} will have variance $1/N$, while the noise has variance~$\nu$ of
order 1. The model we study is related to a family of connectivity matrices
studied by Sompolinsky~\cite{sompolinsky1986neural}, and bears
similarities with a model recently proposed by Mongillo \emph{et
  al.}~\cite{mongillo2018inhibitory}.

In brain networks, connectivity matrices are not symmetric. However, if we assume that the weights depend on
the stimuli only via the symmetric matrix $\mX^* {\left(\mX^*\right)}^\top$, these
asymmetries will be due to the different sources of noise in the learning process, 
and the fact that connectomic reconstructions will give us at best an approximation of true synaptic weights 
(see Discussion). For the purpose of
analysing quantities derived from the symmetric matrix  $\mX^*
{\left(\mX^*\right)}^\top$, as we will do here, we can hence symmetrise the matrix $\mJ$, or equivalently focus on
the case where the noise matrix $\zeta_{ij}$ is symmetric.

\paragraph{A note on the $\nicefrac{1}{\sqrt N}$ scaling} Our choice of scaling
in \cref{eq:w} is made with reconstructing the patterns in mind and follows from
random matrix theory. Our model for the connectivity matrix is related to the
spiked Wigner model~\cite{feral2007largest, potters2020first}, where a random
matrix $\mM$ is constructed as $\emM_{ij} = \beta \evu_i \evu_j + \zeta_{ij}$,
which $\zeta_{ij}$ as above. The matrix~$\mM$ is hence a rank-one perturbation
of a random matrix with elements drawn i.i.d.\ from the normal distribution. The
task is to reconstruct the vector $\vu$, whose elements are of order 1, from the
matrix $\mM$. Intuitively, we want to put ourselves in the ``interesting''
regime, where reconstruction of the $N$ entries of $\vu$ from the
$\mathcal{O}(N^2)$ elements of $\mM$ is neither trivially easy nor impossible. A
classic analysis of this model using random matrix
theory~\cite{feral2007largest} reveals that this ``interesting'' regime is
characterised by a phase transition in the overlap between the leading
eigenvector of $\mM$ and the vector $\vu$, which occurs precisely for a
signal-to-noise ratio $\beta$ that scales as $\nicefrac{1}{\sqrt N}$. We thus
choose the same scaling for our non-linear matrix model~\eqref{eq:J}.

\paragraph{Reconstructing vs retrieving the memories} We emphasise that our
focus in this paper is the \emph{reconstruction} of stimuli from an observed
connectivity matrix $\mJ$, which is different from the \emph{retrieval}
problem~\cite{hopfield1982, amit1985, amit1985a, nishimori2001}, where we ask
whether the patterns $\mX$ are stable fixed points of the dynamics of the
network, \cref{eq:dynamics}. We will discuss the feasibility of reconstructing
patterns from the network's dynamics at the end of the paper.

\subsection*{Solving the inference problem using statistical
  physics}%
\label{sec:inference-problem}%
\addcontentsline{toc}{subsection}{Solving the inference problem using statistical
  physics}

Our aim is to reconstruct the patterns $\mX^*$ that were used to create the
connectivity matrix $\mJ$ of a Hopfield network using the connectivity structure~\eqref{eq:J}. We will call the patterns $\mX^*$ the ground
truth of the problem. Since the connectivity structure is stochastic, we formulate the
pattern reconstruction as a probabilistic inference problem. We interpret the connectivity matrix of the network $\mJ$
as a noisy observation of the symmetric low-rank matrix $\mW \sim \mX^* (\mX^*)^\top$, which was distorted by the transformation given by the rule~\eqref{eq:J}. We can characterise the conditional probability
distribution of a weight $\emJ_{ij}$ given $\emW_{ij}$ as
\begin{subequations}
  \label{eq:channel}
  \begin{align}
    \label{eq:channel1}
    P_\out(\emJ_{ij}=0|\emW_{ij}) &= \frac{1}{2}\erfc\left(\frac{\emW_{ij}-\tau}{\sqrt{2}\nu}\right),  \\
    P_\out(\emJ_{ij}|\emW_{ij}) &= \frac{\exp\left(\nicefrac{-{(\emJ_{ij}-\emW_{ij}+\tau)}^2}{2\nu^2} \right)}{\sqrt{2
                                \pi }\nu} \quad \text{for} \; J_{ij} >  0.\label{eq:channel2}
  \end{align}
\end{subequations}
Reconstructing low-rank matrices from such noisy, distorted observations is a generic inference
problem that appears in a lot of different applications, such as
robust~\cite{Wright2009} and sparse~\cite{Zou2006, Deshpande2014, Lesieur2015b}
PCA, Gaussian mixture clustering~\cite{Matsushita2013}, and community detection
in dense networks~\cite{Fortunato2010a,Decelle2011a}, to name but a
few. Low-rank matrices have also been used extensively in neuroscience to model
recurrent connectivity~\cite{mastrogiuseppe2018linking, schuessler2020interplay,
  sussillo2009generating, logiaco2021thalamic}. The advantage of a Bayesian approach in all these
problems is that they allow for a principled and transparent integration of
knowledge about the problem into the inference process, for example through the
choice of prior distribution and output channel. In the following, we will assume that we know the hyper-parameters $\nu$ and
$\tau$ that were used to generate the connectivity matrix $\mJ$.  Expectation
maximisation and related techniques seem natural candidates to extend our
approach to cases where we would need to learn these hyper-parameters (see
\emph{Discussion}). 

Here, we adopt a Bayesian approach~\cite{mackay2003} to the
inference of the patterns given the connectivity $\mJ$. This means that we will consider our reconstruction of the patterns as a random variable $\mX$, whose posterior distribution $p(\mX | \mJ)$
given the connectivity matrix is given by Bayes' theorem: \begin{equation}
  \label{eq:posterior}
  P(\mX|\mJ) = \frac{1}{Z(\mJ)} \prod_i^N P_X(\vx_i) \prod_{j>i}^N P_{\out}(\emJ_{ij}|
  \emW_{ij}).
\end{equation}
where we introduced the shorthand $\vx_i = \mX_{:,i} \in \mathbb{R}^P$ for the $i$th
column of $\mX$. We  assume that patterns are uncorrelated from
each other and that we know the \emph{a priori} distribution $P_\mX(\mX)$ over
patterns that were stored in the network.
This distribution could for example reflect the fact that we know that the
memories stored in the network are binary, encoding whether a given neuron is
firing or not, or that we have an idea of the probability that any given neuron
is firing in a given memory. 

Note that we are phrasing the problem of reconstructing the memories here as the problem of reconstructing the columns of the matrix $\mX$,
or equivalently, the tuning curves of each neuron. Eventually, we are interested
the whole matrix $\mX$, so it doesn't matter whether we reconstruct all its
columns or all its rows. However, it is more convenient from an algorithmic
point of view to work with the columns, which is the approach we will adopt
here. The marginals of the rows of this distribution, which are $N$-dimensional, provide the best estimate of the patterns that can
be performed~\cite{mackay2003}.

Evaluating the high-dimensional integral to obtain the marginals exactly is an
intractable problem. Instead, here we exploit the formal analogy between the
posterior distribution~\eqref{eq:posterior} and certain probability
distributions that arise in statistical physics to derive an algorithm called
``approximate message passing''~\cite{Donoho2009}, which performs approximate
Bayesian inference of the memories stored in the network. Furthermore, we will
demonstrate that techniques from statistical physics, in particular a tool
called ``state evolution'' (SE), can be used to analyse the behaviour and the
performance of this algorithm in quite some detail. Approximate message passing
algorithms can be understood as a variant of belief propagation, a general
algorithm for inference in graphical models that is usually credited to
Pearl~\cite{Pearl1986}.

\section*{Results}
\addcontentsline{toc}{section}{Results}

\subsection*{A Bayesian algorithm for pattern reconstruction}
\addcontentsline{toc}{subsection}{A Bayesian algorithm for pattern reconstruction}

The inference problem considered here, where we aim to recover a symmetric
low-rank matrix from noisy observations, can be solved using a class of
approximate message passing (AMP) algorithms for low-rank matrix factorisation
called Low-RAMP. It was derived by Lesieur et al.~\cite{Lesieur2017a}, building
on previous works~\cite{fletcher2012,matsushita2013a,Deshpande2014} that
provided AMP algorithms for particular instances of low-rank matrix
factorisation problems. Low-RAMP is an iterative algorithm that produces
estimates for the mean $\hat{\vx}_i$ of the marginal distribution of $p(\vx_i)$ and
their covariance matrix $\sigma_i$, where $\vx_i$ is in general the $i$th column of the low-rank matrix $\mX$ that we are estimating by evaluating the posterior distribution~\eqref{eq:posterior}. In the present case, $\hat{\vx}_i$ is the mean of the estimated `tuning curve' of the $i$th neuron (see above). Using this framework, we will derive
variants of the algorithm for the pattern reconstruction problem outlined in the
previous section. We present the algorithm in detail in the Methods section.

We also provide a
\href{https://github.com/sgoldt/reconstructing_memories}{reference
  implementation} of Low-RAMP for symmetric and
bipartite matrix factorisation problems applicable to a number of different
problems. It is designed to be easily extendable to other problems and also
provides a number of further utility functions. All the results in this paper
can be reproduced using this code.

\subsection*{State evolution}\label{sec:state-evolution}
\addcontentsline{toc}{subsection}{State evolution}

The AMP algorithm has the distinctive advantage over other algorithms, such as
Monte Carlo methods, that its behaviour in the limit $N\to\infty$ for separable prior on the $\mX^*$, random i.i.d. noise $\zeta{ij}$, and number of patterns $P=O(1)$, can be tracked
exactly and at all times using the ``state evolution''
technique~\cite{Donoho2009, Bolthausen2014}. The roots of this method go back to
ideas originally introduced in physics to deal with a class of disordered
systems called glasses~\cite{Mezard1986, Mezard1987}. For the low-rank matrix
factorisation problems we consider here, state evolution was derived and
analysed in detail by Lesieur et al.~\cite{Lesieur2017a}, building on previous
works that derived and analysed state evolution for other specific
problems~\cite{fletcher2012, Deshpande2014, Lesieur2015b}. The last few years in
particular have seen a surge of interest in using state evolution to understand
the properties of approximate Bayesian inference algorithms for a number of
problems~\cite{zdeborova2016}.

Since we are adopting a probabilistic approach to estimating the patterns, we will call the reconstruction of the patterns the mean of the posterior distribution, which we denote by a hat: $\hat{\vx}_i$. Our goal is to track the mean-squared error $\mse_X$ of the
reconstruction $\hat{\vx}_i^t$ of the true signal $\vx_i^*$ after $t$ steps of the
algorithm,
\begin{equation}
  \label{eq:mse}
  \mse(t) \equiv \frac{1}{N} \sum_i^N ||\hat{\vx}^t_i - \vx^*_i||_2^2,
\end{equation}
where $||\cdot||_2^2$ denotes the Euclidean norm of a vector. The $\mse$ can be
expressed in terms of a single matrix-valued parameter defined as
\begin{equation}
  \label{eq:M}
  \mM^t \equiv \frac{1}{N} \sum_i^N \hat{\vx}_i^t \vx_i^{*, \top} \in \mathbb{R}^{P
    \times P},
\end{equation}
such that
$\mse(t) = \Tr\left[ \langle \vx_0 \vx_0^\top \rangle -
  \mM^t\right]$. Here and throughout this paper we write averages with respect
to the prior distribution $p_X(\vx)$ of the corresponding model as
$\langle \cdot \rangle$. We write $\vx_0$ with the subscript to underline that the random variable $\vx_0$ is not a column of the matrix $\mX$ that we're trying to evaluate; instead, it is a variable that is drawn from the prior and averaged over. 

Now the goal is to find an update equation for the order parameter $M^t$ that
mirrors the progress of the algorithm. This update equation is the state
evolution equation~\cite{Donoho2009, Bolthausen2014}. Remarkably, from~\cite{lesieur2016b} we see
that the two constants defining our problem, $\tau$ and $\nu$, do not appear
explicitly in the state evolution equations. Instead, the behaviour of the
algorithm -- and hence its performance -- only depends on an effective
signal-to-noise ratio (SNR) of the problem, which is a function of the
threshold~$\tau$ and noise variance~$\nu$ utilised in the connectivity structure~\eqref{eq:J}. Formally, it can be expressed as the inverse of
the Fisher score matrix~\cite{cover2006a} of the generative model we use to
describe how the network is connected~\eqref{eq:channel1}-\eqref{eq:channel2}, evaluated
at~$\emW_{ij}=\vx_i \vx_j^\top /\sqrt{N} = 0$:
\begin{align}
  \label{eq:Delta}
  \frac{1}{\Delta} & \equiv \mathbb{E}_{P_\mathrm{out}(J|w=0)} {\left(
                     \frac{\partial \ln P_{\mathrm{out}}(J|w)}{\partial w} \right)}^2_{J,w=0} \\
                   & = \frac{\tau e^{-\nicefrac{\tau^2}{2\nu^2}}}{\sqrt{2 \pi}
                     \nu^3} + \frac{e^{-\nicefrac{\tau^2}{\nu^2}}}{\pi \nu^2
                     \erfc\left(\nicefrac{-\tau}{\sqrt{2}\nu}\right)} + \frac{1}{2\nu^2}\erfc\left(\frac{\tau}{\sqrt{2}\nu} \right).
\end{align}
Here and throughout, $\mathbb{E}$ denotes the expectation over the random
variables. In fact, on the level of the algorithm, everything about the output
channel~\eqref{eq:channel} can be summarised in this single, scalar quantity
$\Delta$. This remarkable universality of the state evolution and hence the AMP
algorithm with respect to the output channel was first observed
in~\cite{lesieur2016b} and dubbed ``channel universality''.

State evolution provides an update equation for the order parameter $M^t$ that
mirrors the progress of the algorithm. We first define an auxiliary function
\begin{equation}
  \label{eq:fin}
  f(\mA, \vb) =  \frac{1}{Z(\mA, \vb)} \sum_{\vx\in\mathcal{X}^P} \vx\, p_X(\vx) \exp \left( \vb \vx -
    \frac{1}{2}\vx^\top \mA \vx \right).
\end{equation}
where $\mA\in\mathbb{R}^{P\times P}$ and $\vb\in\mathbb{R}^P$. If $\mA=0$ and
$\vb=0$, this function would compute the average over the prior distribution
$p_X(\vx)$. Instead, $\vb$ and $\mA$ are estimated from the data (see the
algorithm for details) so $f$ computes an average over a distribution that
contains the prior and a data-dependent part. This structure reflects the
Gaussian approximation of the posterior density that we apply here, or more
broadly speaking the interplay between prior information and data-dependent
likelihood that is typical of Bayesian inference. Consequently,
$Z(\mA, \vb) = \sum_{\vx\in\mathcal{X}^P} p_X(\vx) \exp \left( \vb \vx -
  \nicefrac{1}{2}\, \vx^\top \mA \vx \right)$ is a normalisation factor. The
update equation for the order parameter $M^t$ can be written using this
auxiliary function for all the cases considered in this paper; it
reads~\cite{Lesieur2017a}
\begin{equation}
  \label{eq:se}
  \mM^{t+1} = \underset{\vx_0, \vz}{\mathbb{E}} \left[
    f\left(\frac{\mM^t}{\Delta}, \frac{\mM^t}{\Delta} \vx_0 +
      \sqrt{\frac{\mM^t}{\Delta}}\vz\right) \vx_0^\top \right]
\end{equation}
where $\vz$ is a $P$-dimensional vector of Gaussian random variables with mean
zero and variance~1. The average over $\vx_0$ is taken with respect to the prior
distribution $p_X(\vx)$, as discussed above.

So to summarise, statistical physics gives us an algorithm to perform
approximate inference of the patterns and the state evolution
equation~\eqref{eq:se} allows us to track the behaviour of the algorithm over
time. We can thus analyse the performance of the algorithm in high-dimensional
inference by studying the fixed points of the low-dimensional
state-evolution~\eqref{eq:se}. This is the key idea behind this approach, and we
will now demonstrate the usefulness of this machinery by applying it to several
specific cases.

\subsection*{Reconstructing binary patterns}
\addcontentsline{toc}{subsection}{Reconstructing binary patterns}

As a first application of the algorithm and the analysis tools outlined so far,
we consider the reconstruction of a set of binary patterns,
$\mathcal{X}=\{\pm1\}$. We will assume that both positive and negative values
are equiprobable and that the components of a pattern vector are independent of
each other, so the prior on a column of the matrix of stored patterns, $\vx_i$, is simply
\begin{equation}
  \label{eq:prior_hf}
  p_X(\vx_i) = \prod_j^P p_x(\emX_{ij}) = \frac{1}{2^P}.
\end{equation}

\subsubsection*{A single pattern ($P=1$)}
\addcontentsline{toc}{subsubsection}{A single pattern}

It is instructive and helpful for the following discussions to first consider
the case where $P=1$, \emph{i.e.}\ there is only a single pattern stored in the
network that we are trying to recover from $\mJ$. The threshold function for the
model then becomes $f(A, B) = \tanh(B)$, with $B \in \mathbb{R}$,
and the state evolution for the now scalar parameter $m^t$ simplifies to
\begin{equation}
  \label{eq:se-P1}
  m^{t+1} = \underset{z}{\mathbb{E}} \; \tanh\left(\frac{m^t}{\Delta} 
    + \sqrt{\frac{m^t}{\Delta}}z\right)
\end{equation}
where  $w$ is a scalar Gaussian random variable with zero mean and unit
variance. 

\begin{figure}
  \centering
  \includegraphics[width=.38\textwidth]{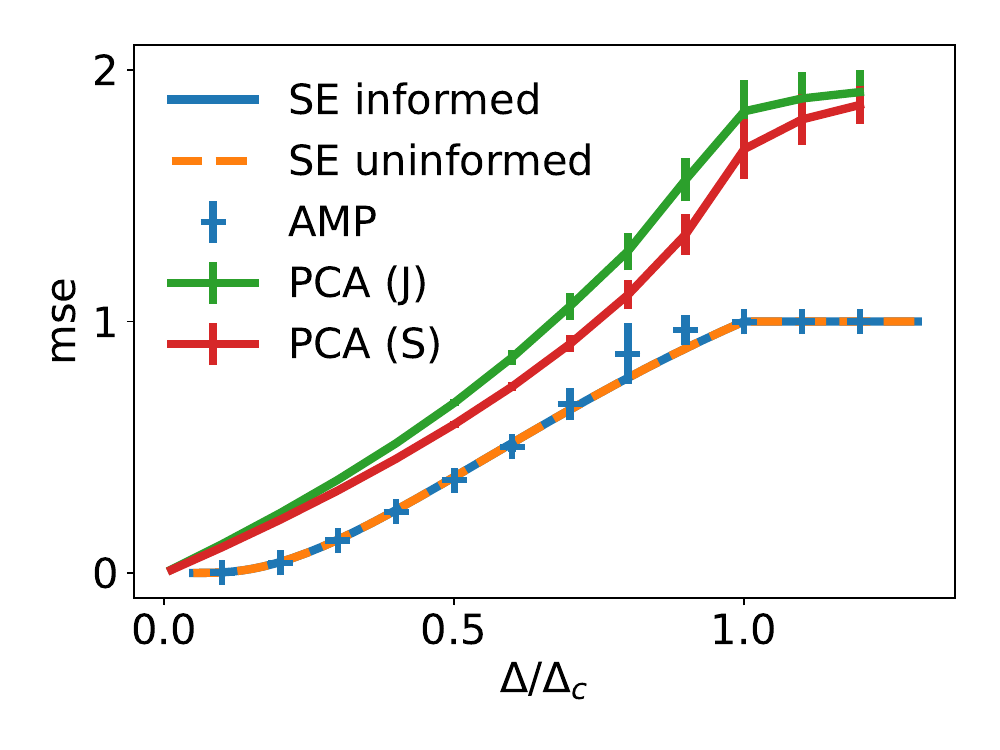}%
  \includegraphics[width=.29\textwidth]{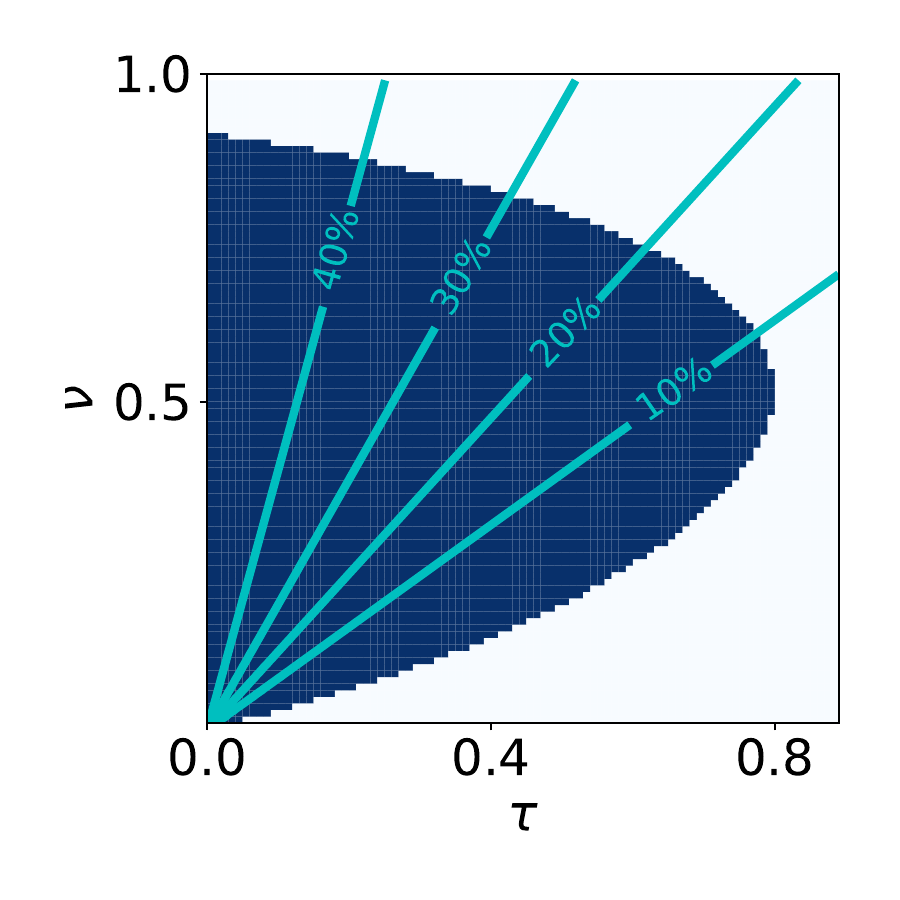}%
  \includegraphics[width=.37\textwidth]{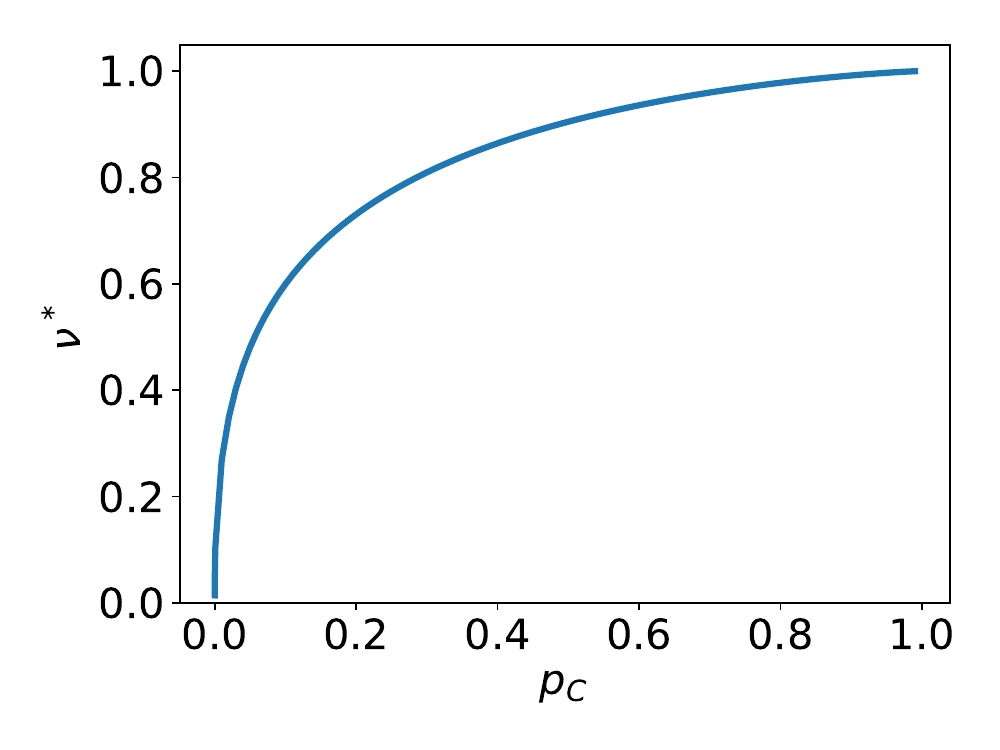}
  \caption{\label{fig:amp-vs-se_hf} \figleft{} \textbf{Reconstruction and
      performance of the message-passing algorithm for binary patterns.} We plot
    the $\mse$~\eqref{eq:mse} obtained by the AMP algorithm~\eqref{eq:amp_tap}
    as a function of the effective noise $\Delta$~\eqref{eq:Delta} (blue
    crosses). We plot the performance of the algorithm starting from
    random~\eqref{eq:random} and informed~\eqref{eq:informed}
    initialisations. Solid lines depict the prediction obtained from iterating
    the state evolution equation~\eqref{eq:se-P1}. Having $\Delta/\Delta_c > 1$
    corresponds to the white region in the phase diagram on the right. We also
    plot the $\mse$ of the reconstruction obtained by applying PCA to the weight
    matrix $\mJ$ and to the Fisher matrix $\mS$~\eqref{eq:S} (green
    and red, resp.) \emph{Parameters:} $\tau=0$. $N=5000$ for AMP, $N=20000$ for
    PCA. \quad \figcenter{} \textbf{Phase diagram for the rectified Hopfield
      channel with $P=1$.} We plot whether reconstruction of the patterns better
    than a random guess is easy (blue) or impossible (white) using the
    message-passing algorithm as a function of the constant threshold $\tau$ and
    the variance $\nu$ of the Gaussian noise appearing in the connectivity
    structure~\eqref{eq:J}. The solid lines are the contours of the
    connection probability $p_C(\nu, \tau)$~\eqref{eq:p_C}. \quad \figright{}
    \textbf{Critical noise $\nu^*$ as a function of connection probability
      $p_C$.} We plot $\nu^*$, the largest variance of the additive Gaussian
    noise $\zeta_{ij}$ at which reconstruction remains possible, against the
    probability $p_C$~\eqref{eq:p_C} that any two neurons are connected.}
\end{figure}

We can now iterate the state evolution equation~\eqref{eq:se-P1} with a given
noise level $\Delta(\nu, \tau)$ until convergence and then compute the $\mse$
corresponding to that fixed point. The fixed point we converge to reveals
information about the performance of the AMP algorithm. We plot the results on
the left-hand side of \cref{fig:amp-vs-se_hf} for the two different
\emph{initialisations} of the algorithm: in blue, we plot the $\mse$ obtained by iterating SE starting with an
\emph{random initialisation}
\begin{equation}
  \label{eq:random}
  m^{t=0}=0 + \delta,
\end{equation}
where $\delta>0$ is a
very small random number. The error obtained in this way is the one that is
obtained by the AMP algorithm when initialised with a random draw from the prior
distribution -- in other words, a random guess for the patterns. This is confirmed by the blue crosses, which show the mean and
standard deviation of the $\mse$ obtained from five independent runs of the
algorithm on actual instances of the problem. The dashed orange line in \cref{fig:amp-vs-se_hf} shows the final $\mse$
obtained from an informed initialisation
\begin{equation}
  \label{eq:informed}
  m^{t=0}=1-\delta,
\end{equation}
which would
correspond to initialising the algorithm with the solution,
i.e. $\hat{x}_i=\vx_i^*$.

In this model, we find that the AMP algorithm starting from a random guess
performs just as well as the algorithm starting from the informed
initialisation. This need not always be the case, and we will indeed find a
different behaviour in the next sparse and skewed models we consider. 

\paragraph*{When is recovery possible?}

We can see from the middle plot of \cref{fig:amp-vs-se_hf} that recovery
of the memories from the connectivity $\mJ$ is not always possible; there exists
a critical value for the effective noise $\Delta_c$ above which the mean-squared error of
the solution obtained by the algorithm is the same as we would have obtained by  making a random guess for
the solution based on the prior distribution~\eqref{eq:prior_hf} alone, without
looking at the data. We can
calculate this critical noise level $\Delta_c$ using the state
evolution~\eqref{eq:se}. We can see from that equation that $m^t=0$ is a trivial
fixed point, in the sense that the $\mse$ corresponding to that fixed point is
equal to the $\mse$ obtained by making a random guess. Expanding
\cref{eq:se-P1} around this fixed point yields $m^{t+1} = m^t/\Delta$.
There are hence two regimes for recovery, separated by a critical value
\begin{equation}
  \label{eq:4}
  \Delta_c=1
\end{equation}
of the effective noise~\eqref{eq:Delta}. If $\Delta > \Delta_c$, the uniform
fixed point is stable and recovery is impossible. On the other hand, for
$\Delta < \Delta_c$, the uniform fixed point is unstable and hence AMP
returns an estimate for the patterns that has an $\mse$ that is lower than
random guessing. The phase diagram in the middle of \cref{fig:amp-vs-se_hf}
delineates the easy and the impossible phase for the rectified Hopfield channel
with symmetric prior~\eqref{eq:prior_hf}. While there could be in principle other fixed points of the
state evolution equations for other priors and channels~\cite{Lesieur2017a}, it is always one of the fixed
points that is reached from either the informed or the uninformed initialisation
that describes the behaviour of the algorithm.


At first sight, the impact of the additive Gaussian noise $\zeta_{ij}$ on the
phase diagram in \cref{fig:amp-vs-se_hf} appears counter-intuitive. If we
fix the threshold to, say, $\tau=0.5$, reconstruction is impossible for small
variances $\nu$ of $\zeta_{ij}$. As we increase $\nu$, i.e.\ as we add
\emph{more} noise to the system, recovery becomes possible.  The key to
understanding this behaviour is that for a single stimulus $P=1$, a weight in
the network will have one of two possible values which are symmetric around the
origin, $\emW_{ij}=\pm \nicefrac{1}{\sqrt N}$. By applying the rectification, for any cut-off $\tau>\emW_{ij}$
the resulting weight matrix $\mJ$ without additive noise is trivially zero and no
recovery is possible. We can only hope to detect something when an added noise $\zeta_{ij}$
pushes the value of the weight before rectification above the cut-off. Recovery
then becomes possible if the added noise is large enough that the weight without
noise is larger than the cut-off $a + \zeta_{ij} > \tau$, while remaining small
enough that it's significantly more likely that the noise-less weight is
positive than negative. As the noise variance increases even further, its
detrimental effects dominate, and recovery becomes impossible again. This
mechanism is reminiscent of stochastic resonance (SR), a mechanism where a weak
signal is amplified by the presence of noise. Indeed, our problem contains the three ingredients for SR
(e.g.~\cite{gammaitoni1998stochastic}): A
threshold mechanism, given by the rectification in the connectivity structure: A weak signal (the
stored patterns); and a noise term, $\zeta$.

As already mentioned, when noise is too large recovery becomes impossible. We show on the right of
\cref{fig:amp-vs-se_hf} the critical variance of
$\zeta_{ij}$ above which reconstruction becomes impossible, $\nu^*$, as a function of
the connection probability $p_C$, given by Eq.~\eqref{eq:p_C}. This plot can be obtained by
solving, for a given value $p_C=c$, the two-dimensional system
\begin{align}
  \Delta &= 1 \\
  p_C &= c
\end{align}
for $(\tau, \nu)$. As expected, the critical variance increases with
the connection probability, and it goes to zero as the connection
probability goes to zero.

\subsubsection*{Comparison to principal component analysis (PCA)}
\addcontentsline{toc}{subsubsection}{Comparison to principal component analysis (PCA)}


Principal component analysis (PCA) is another method to reconstruct the stored
patterns from the network connectivity. PCA and other spectral methods have some
advantages: they are non-parametric, and their implementation in the case of a
single pattern is straightforward: the PCA prediction for the stored pattern is
simply the leading eigenvector of $\mJ$. We plot the mean-squared
error~\eqref{eq:mse} of this estimate with the green line on the left of
\cref{fig:amp-vs-se_hf}, where we see that the reconstructing error of PCA is
larger than the one of AMP, especially for large values of the
noise. This is also borne out by theory: the reconstruction mean-squared error
of PCA can be shown to be strictly larger than the AMP
estimate, since the latter is the Bayes-optimal predictor~\cite{Lesieur2017a}.

An alternative PCA algorithm can be found by linearising the AMP equations
around the trivial fixed point $\hat \vx=0$~\cite{krzakala2013spectral,
  zdeborova2016}. This linearisation yields an equation that can be interpreted
as PCA applied to the Fisher matrix $\mS$~\eqref{eq:S} instead of~$\mJ$. Since
the Fisher matrix depends on the generative model for the data when deriving the
message-passing equations, looking at its leading eigenvector offers a spectral
algorithm that is more adapted to the problem at hand. Indeed, we find that its
error (red line in \cref{fig:amp-vs-se_hf}) is slightly lower than the error
obtained from PCA on the weights directly. In either case, the performance of
PCA is worse than that of AMP.

The large value of the PCA error compared to the AMP error at large noise levels
in \cref{fig:amp-vs-se_hf} reveals a fundamental weakness of PCA: even at noise
levels above the critical noise $\Delta_c$, where no reconstruction is possible
for any algorithm, PCA can be applied and will return a prediction -- there is
no concept of uncertainty in PCA. Hence the $\mse$ of PCA tends to a constant as the
noise increases and the leading eigenvector of $\mJ$ is just a random
vector; for the Hopfield prior and when rescaling the eigenvectors to have the same length as draws from the Hopfield prior, this constant is 2. AMP on the other hand returns a vector full of zeros if
$\Delta > \Delta_c$ (and the prior has an average of 0, as is the case for all
the priors we consider). AMP thus expresses its uncertainty about the planted
pattern, yielding an $\mse = 1$ for inputs with~$x_i=\pm1$. The advantage of the
Bayesian approach is thus that it prevents over-confident predictions in the
high noise regime.



The weaker performance of PCA compared to AMP is due to the fact that spectral methods do not not offer a
natural way to incorporate the prior knowledge we have about the structure and
distribution of stimuli into the recovery algorithm. The Bayesian framework
incorporates this domain knowledge in a transparent way through the generative
model of the stored patterns~$p_X(\vx)$. We will see that this creates an even
larger performance gap for sparse patterns and patterns with low coding level.


\subsubsection*{Many patterns ($P>1$)}%
\addcontentsline{toc}{subsubsection}{Many patterns ($P>1$)}%
\label{sec:many-patterns}

For the general case of several patterns $P>1$ with finite $P$, we can
significantly simplify the state evolution by noticing that the matrix $M^t$
will interpolate between a matrix full of zeros at time $t=0$ and a suitably
scaled identity matrix in the case of perfect recovery, \emph{i.e.}
\begin{equation}
    \label{eq:dfn-a}
    M^t = m^t \mI_P / \Delta,
\end{equation}
where $\mI_P$ is the identity matrix in $P$
dimensions. In other words, for uncorrelated patterns, the different input
patterns do not interact during the reconstruction, and so the off-diagonal
matrix elements remain zero in the case where we only store a few patterns and
the connectivity structure remains low-rank. Once we overload the model by
storing many more patterns, we would have non-zero off-diagonal elements,
meaning that reconstructions converge to spurious patterns, for example linear
combinations of the patterns. However, in this regime the
state evolution derived above also breaks down.
In this case, the threshold function becomes
\begin{equation}
  [{f(A, B)}_k] =  \tanh\left( \frac{m^t}{\Delta}x_{0,k} + \sqrt{\frac{m^t}{\Delta}}z_k\right)
\end{equation}
where $z_k$ is again a standard Gaussian variable. Substituting into the state evolution gives
an update equation for the parameter $m^t$, namely
\begin{equation}
  \label{eq:hf_se}
  m^{t+1} = \underset{z}{\mathbb{E}} \;  \tanh\left(
    \frac{m^t}{\Delta} + \sqrt{\frac{m^t}{\Delta}}z\right)
\end{equation}
where $m^t$ is the overlap parameter introduced above~\eqref{eq:dfn-a}. This update has the same form as the state evolution in the $P=1$ case,
\cref{eq:se-P1}. So we find, remarkably, that recovering $P$ distinct patterns
is exactly equivalent to recovering a single pattern $P$ times in the
thermodynamic limit where $N\to\infty$ while the number of patterns is of order
$P\sim\mathcal{O}(1)$. This approximation will eventually break down in
practical applications with finite network sizes, and we investigate the
breaking point of this behaviour below.

Recovering many patterns with PCA poses an additional challenge. While it is
easy to recover the leading rank-$P$ subspace of the matrices $\mJ$ or $\mS$ by
simply computing the $P$ leading eigenvectors, it is not clear how to recover
the exact patterns from these eigenvectors, which can be any rotation of the
input patterns due to the rotational symmetry of $\mW$. This can be seen from
the fact that the patterns $\mX^*$ could be multiplied by any rotational matrix
$\mO$ with $\mO \mO^\top=\mathbb{I}$ without changing the resulting weight
matrix $\mJ$, see \cref{eq:w}. The best way to recover the exact stimuli from
the principal components is thus not clear \emph{a priori} (see \cite{blumenfeld2010algorithm}). 
Other problems require combining PCA with other methods, such as $k$-means or
gradient descent. Since we have already seen that AMP outperforms PCA on binary
patterns, and we will see that this gap only increase for the other types of
patterns we will study below, we do not investigate further this direction.

\subsubsection*{A first summary} 
\addcontentsline{toc}{subsubsection}{A first summary} 

Before we turn to more complicated prior distributions over the patterns, let us
briefly summarise our results so far. We derived an algorithm that can
reconstruct patterns from the connectivity of recurrent network whose weights
are obtained from the learning rule \cref{eq:J}. Whether or not the algorithm is
successful in this reconstruction depends on the noise level $\nu$ and the
threshold $\tau$. These parameters can be combined into an effective noise
parameter $\Delta$~\eqref{eq:Delta}, which determines the performance of the
message-passing algorithm. The algorithm performs well in the reconstruction
task, and beats non-parametric approaches like PCA by virtue of including prior
information about the distribution of the patterns in a principled way.

\subsection*{Sparse patterns}%
\addcontentsline{toc}{subsection}{Sparse patterns}%
\label{sec:sparse-rect-hopf}

An interesting variation of the rectified Hopfield model is its sparse version,
where only a fraction $0\le\rho\le1$ of the components $x_{ij}$ of a pattern
$\vx_i$ are non-zero. The prior distribution then becomes
\begin{equation}
  \label{eq:prior_hf-sparse}
  p_X(\vx_i) = \prod_{j=1}^P p_x(x_{ij}) = \prod_{j=1}^P\left[ (1-\rho)\delta(x_{ij}) + \nicefrac{\rho}{2}\left[ 
      \delta(x_{ij} - 1) + \delta(x_{ij} + 1)\right]\right],
\end{equation}
where $\delta(\cdot)$ is the Kronecker delta. This prior has mean
$\langle x \rangle=0_p$, where $0_p$ is a vector of $p$ zeros, and covariance
$\langle x x^\top \rangle=\rho \mI_p$. We emphasize again that these patterns represent a deviation of neuronal activity from its mean - in this case $\pm 1$ represents an increase/decrease activity, while 0 represents no change in activity in a given pattern. The state evolution will interpolate
between an order parameter that is all zeros for an estimator that is drawn from
the prior distribution and completely uncorrelated with the ground truth, and
$M=\rho \mI_P$ for perfect reconstruction. We delegate the mathematical details
to the Methods section, and focus here on the performance of the algorithm.

\begin{figure}[t!]
  \centering 
  \includegraphics[width=\linewidth]{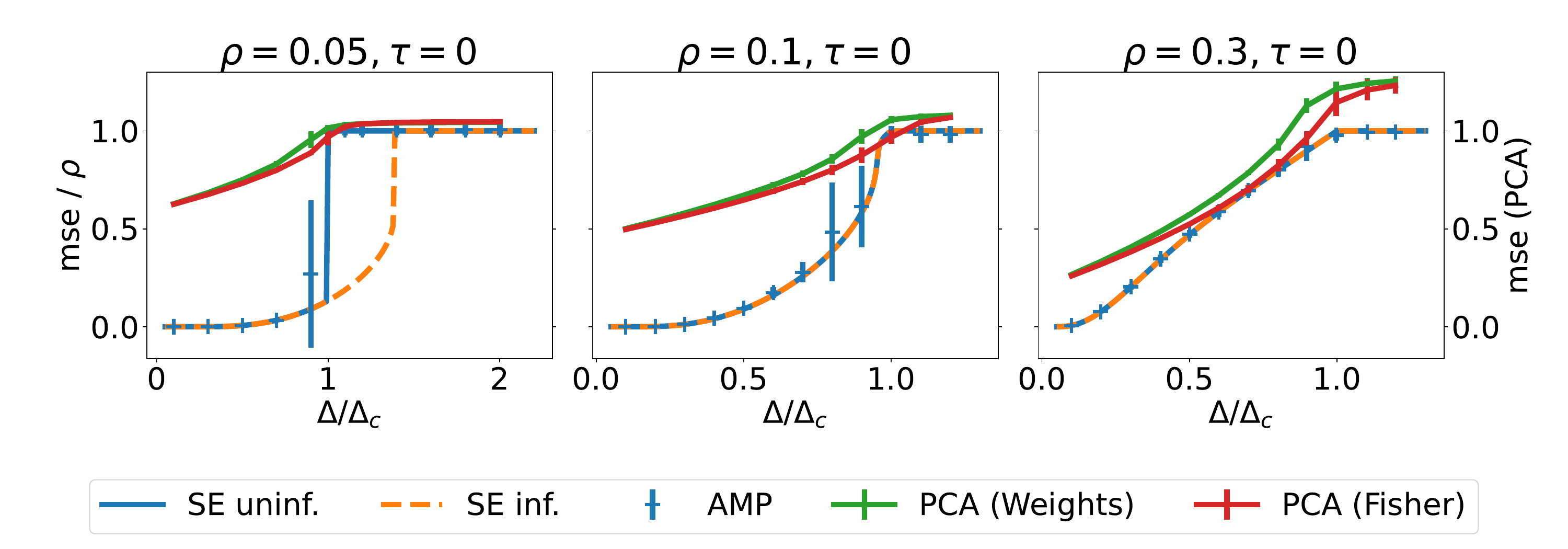}
  \caption{\label{fig:amp-vs-se_hf-sparse} \textbf{Reconstructing \emph{sparse},
      binary patterns using message passing algorithms and PCA.} We plot the
    $\mse$ per pattern obtained by the AMP algorithm, \cref{eq:amp_tap}, as
    a function of the effective noise $\Delta$~\eqref{eq:Delta}, for
    random~\eqref{eq:random} and informed~\eqref{eq:informed}
    initialisations. Lines depict the result of the state evolution, while
    crosses denote the performance of the AMP algorithm on an instance of the
    problem. While AMP performs the same starting from both initialisations for
    $\rho=0.1$ and $\rho=0.3$, there is a gap in performance for $\rho=0.05$,
    which might hint at the existence of a hard phase (see main text). We also
    plot the $\mse$ of the reconstruction obtained by applying PCA to the weight
    matrix $\mJ$ and to the Fisher matrix $\mS$~\eqref{eq:S}
    (green and red, resp.) \emph{Parameters:} $\tau=0$. $N=2000$ for AMP,
    $N=20000$ for PCA.}
\end{figure}

\paragraph*{Performance of message-passing} We first note that the critical
noise level above which AMP will not be able to recover the stored pattern
better than a random guess in the sparse reconstruction problem is
$\Delta_c=\rho^2$.  In that sense, the critical noise is a property of the AMP
algorithm, and its value can again be obtained by linearising the state
evolution equation~\eqref{eq:rhf_se} around its trivial fixed point
$m^t=0$.  
However, it is generally believed that no other algorithm is able to recover the
patterns above this noise level, which would make this threshold a property of
the model rather than the algorithm; we'll come back to this point in the next
paragraph. In any case, the decrease of $\Delta_c$ with $\rho$ means that the
reconstruction performance at fixed $\nu$, $\tau$ and hence $\Delta$ decreases
with pattern sparsity. However, so too does the difficulty of the problem. If we
normalise the noise level by the critical noise, \emph{i.e.}  if we plot the
reconstruction error as a function of the $\Delta / \Delta_c$ as we do in
\cref{fig:amp-vs-se_hf-sparse}, we see that a large fraction~$\rho$ of non-zero components $\emX_{ij}$ leads to better reconstruction at small noise levels.

For $\rho=0.1$ and $\rho=0.3$, the performance plots in
\cref{fig:amp-vs-se_hf-sparse} resemble the results obtained for the symmetric
Hopfield model overall: Reconstruction is possible if the effective noise level
is below the critical noise level. For these two values of the sparsity, the
$\mse$ of the estimate returned by AMP (blue dashed line) matches the $\mse$
obtained from state evolution when starting from the informed initialisation
(orange). However, the reconstruction errors predicted by state evolution for
informed and random initial conditions \emph{disagree} at smaller sparsity. For
$\rho=0.05$ (left), there is a range of effective noise levels $\Delta$ for
which the performance of AMP does not match the performance predicted by state
evolution starting from informed initial conditions. We note that this could be
the signature of a so-called \emph{hard phase}, where a better-than-chance
reconstruction of the pattern is information-theoretically possible --
i.e.~there is some trace of the stored pattern in the connectivity -- but AMP is
not able to extract it. However, it is important to emphasise that AMP performs sub-optimally with respect to the
amount of information that is in the connectivity, but not with respect to the performance of any other known algorithm. In other words, while AMP does not exploit all the information that is in the connectivity, it is broadly believed in theoretical computer science that
\emph{no algorithm} can reconstruct the patterns with non-trivial error at this
level of noise in polynomial time, if this is indeed a hard phase of the reconstruction problem. PCA for example is also not able to
reconstruct any trace of the pattern at this noise level (see below). We refer
the interested reader to Refs.~\cite{zdeborova2016, bandeira2018notes} for
recent reviews on the topic, with many other examples of problems that exhibit
such a hard phase or computational-to-statistical gap, as it is also sometimes
called in the literature.


\paragraph{Performance of PCA} \Cref{fig:amp-vs-se_hf-sparse} also shows the
reconstruction error of PCA applied to either the weights (green) or the Fisher
matrix (red). Note that while the error of AMP is rescaled by $\rho$, we plot
the reconstruction of PCA without rescaling (which is why we have a second
$y$-axis in all three plots). In other words, the reconstruction error of AMP is
multiplied by 20 in the left-most plot of \cref{fig:amp-vs-se_hf-sparse}, and is
still below that of PCA. We rescale the AMP error in this way to ensure that the
errors are comparable for different values of the sparsity, since AMP returns a
sparse estimate in all cases. PCA on the other hand is agnostic about the
sparsity of the patterns, and returns a dense reconstruction regardless of the
value of $\rho$. The PCA error at high noise thus scales as $1 + \rho$. Again,
we see that AMP outperforms PCA in terms of the reconstruction error, with the
largest difference coming at lowest sparsity. This is to be expected, since the
AMP algorithm can take information about the prior into account. 

\subsection*{Reconstructing patterns with low coding level}
\addcontentsline{toc}{subsection}{Reconstructing patterns with low coding level}

As a third and final example, we consider the reconstruction of patterns with
low coding level. For this, we will draw patterns from the prior distribution
\begin{equation}
  \label{eq:prior_tsodyks}
  p_X(\vx_i) =\prod_{j=1}^P p_x(x_{ij}) = \prod_{j=1}^P\left[ (1-\rho)\delta(x_{ij}+\rho) + \rho \delta(x_{ij} - (1-\rho))\right]
\end{equation}
which is related to models proposed first by Tsodyks and
Feigel'man~\cite{tsodyks1988a, tsodyks1988b}, that considered the storage of binary (0,1) patterns of activity (where 0 means inactive and 1 active), with a `coding level' $\rho$ (probability that a neuron is active in a  given pattern). The motivation for studying the sparse $\rho\ll 1$ limit is that the activity in brain structures involved in memory is typically sparse - for instance, $\rho$ has been estimated to be on the order of 0.01 in animals ranging from rodents to humans \cite{lee20,waydo06}. As in the previous cases, the patterns correspond to deviations of activity from its mean, i.e. ($1-\rho$,$-\rho$) instead of (1,0). This model has mean zero and a
covariance matrix $\rho(1-\rho)\mI_P$. We still use
the channel corresponding to the connectivity structure of the rectified
Hopfield model~\eqref{eq:J}, so while the Fisher score matrix $\mS$ stays the
same, we have a new threshold function and hence a new state evolution, which we
derive in \emph{Methods}.

\begin{figure}[t!]
  \centering
  \includegraphics[width=.8\textwidth]{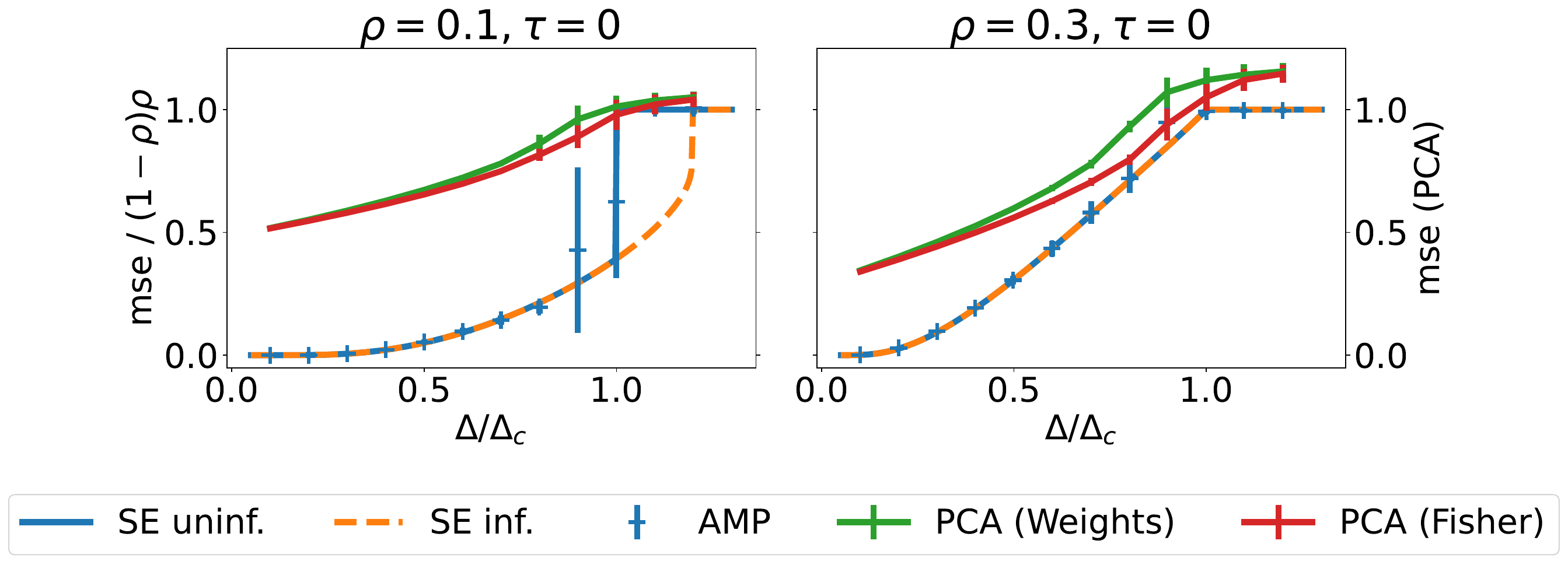}\\
  \caption{\label{fig:amp-vs-se_tsodyks} \textbf{Performance of various
      algorithms when reconstructing a pattern with low coding level following
      Tsodyks' prior.}  \figtop~Same plot as \cref{fig:amp-vs-se_hf}
    and~\ref{fig:amp-vs-se_hf-sparse}, for a single stored pattern drawn from
    Tsodyks' prior~\eqref{eq:prior_tsodyks}. For $\rho<0.211$, the performance
    of the algorithm with random initialisation~\eqref{eq:random} is different
    from the performance with informed initialisation~\eqref{eq:informed}, which
    might be the signature of a hard phase. 
    \emph{Parameters:} $\tau=0$, $N=5000$ for AMP, $N=20000$ for PCA.}
\end{figure}

Here, we plot the performance of the algorithm together with the results of
iterating the state evolution equation (\emph{Methods}) for two different values
of the coding level $\rho$ in \cref{fig:amp-vs-se_tsodyks}. Note that we again
rescale the reconstruction error of AMP by $(1-\rho)\rho$ to ensure
comparability of the results for different values of the coding level
$\rho$. The error curves for PCA (green and red) in \cref{fig:amp-vs-se_tsodyks}
on the other hand are not rescaled. We note that again, AMP outperforms PCA in
terms of the reconstruction error, which for random guessing tends to
$1 + \rho(1-\rho)$ in the case of PCA. 

\paragraph*{The hardness of recovering patterns with low coding level} In this
model too, we have a uniform fixed point with $m^t=0$. Expanding around this
fixed point yields the update equation
\begin{equation}
  m^{t+1} = \frac{{(\rho -1)}^2 \rho ^2}{\Delta }m^t
\end{equation}
so we find that the critical value of the noise where the uniform or
uninformative fixed point becomes unstable in this model is
$\Delta < \Delta_c = {(\rho -1)}^2 \rho ^2$.  

Recently, a closed-form sufficient criterion for the existence of a hard phase
in models that have a prior with zero mean was derived in~\cite{Lesieur2017a};
namely, a hard phase exists if the prior is ``skewed'' in the statistical sense,
such that
\begin{equation}
  \label{eq:order-criterion}
  {\langle x^3\rangle}^2 > 2 {\langle x^2 \rangle}^3
\end{equation}
where the average is taken with respect to the prior distribution of this model,
\cref{eq:prior_hf-sparse}. For the prior~\eqref{eq:prior_tsodyks}, this
criterion predicts the existence of a first order phase transition and hence of
a hard phase for $\rho > \rho_c = 1/2 - 1/\sqrt{12} \simeq 0.2113$ where we
assume w.l.o.g.\ that $\rho < 1/2$. Note that this is a sufficient condition,
and not a necessary one; in fact, for the sparse Hopfield prior, which is
symmetric around zero and has ${\langle x^3\rangle}=0$, we cannot
calculate the critical value of $\rho$ using \cref{eq:order-criterion}.

\subsection*{Reconstructing even more patterns: how far can we go?}
\addcontentsline{toc}{subsection}{Reconstructing even more patterns: how far can we
  go?}%
\label{sec:meanfield-prior}

A natural question that arises for the algorithms we have derived is how many
patterns we can reliably reconstruct.  In practice, the bottleneck for
reconstructing the patterns using AMP is computing the 
partition function of the Gaussian
approximation of the posterior density of a column-vector $\vx$ at every step of the algorithm (see the
detailed explanation of the AMP algorithm in the Methods section):
\begin{equation}
  \label{eq:f_in_sum}
  W(\vx; \mA, \vb) =  \frac{1}{Z(\mA, \vb)} p_X(\vx) \exp \left( \sum_j^P \evb_j \evx_j -\frac{1}{2} \sum_{j,k}^P
    \evx_j \emA_{jk} \evx_k \right).
\end{equation}
Evaluating the mean and the variance of this distribution even for the simple Hopfield prior~\eqref{eq:prior_hf}
requires summing  $2^P$ terms, so the computational cost is exponential in
the number of patterns stored. We can circumvent this bottleneck by computing
the function $W(\vx; \mA, \vb)$ using a mean-field approximation~\cite{Barber2012}, which was
originally proposed in~\cite{manoel2017}.

\begin{figure}[t!]
  \centering
  \includegraphics[width=.33\textwidth]{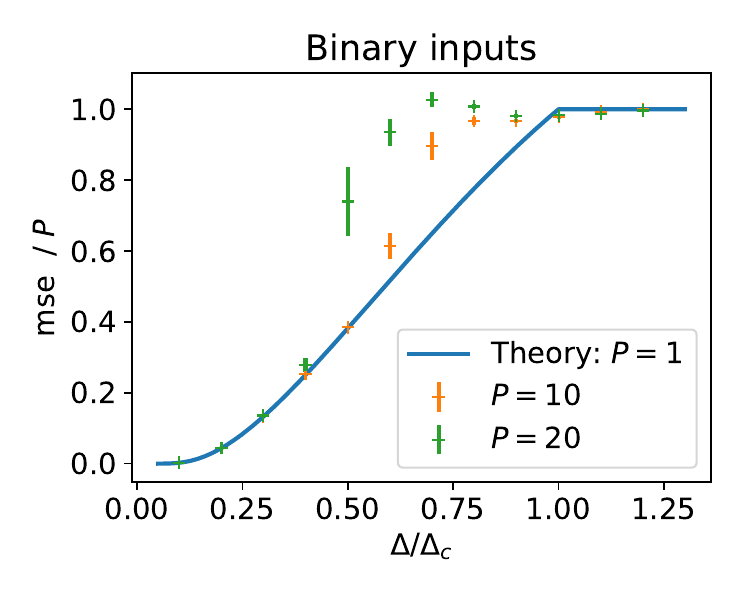}%
  \includegraphics[width=.33\textwidth]{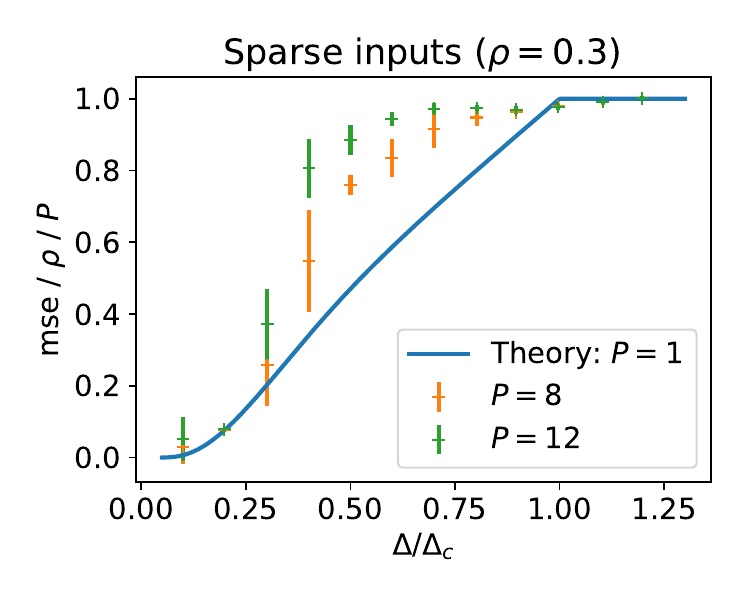}%
  \includegraphics[width=.33\textwidth]{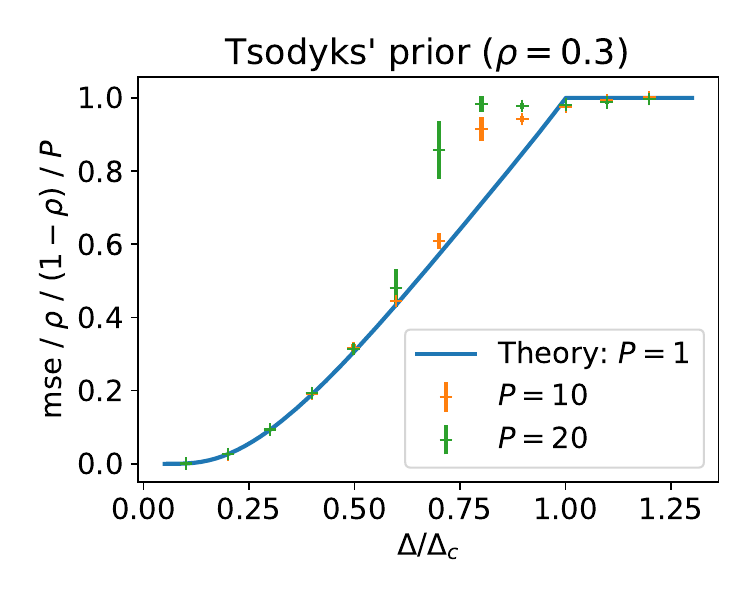}%
  \caption{\label{fig:amp-vs-se_mf} \textbf{MSE from reconstructing with the
      mean-field prior approximation.} We plot the final $\mse$ obtained by the AMP
    algorithm (no damping), using a mean-field approximation to compute the
    threshold function. The solid line is the $\mse$ predicted by iterating the
    state evolution for the scalar variable, \cref{eq:se-P1}. We choose
    $N=5000$ and set $\tau=0$.}
\end{figure}

\paragraph{The mean-field approximation} We thus approximate the posterior distribution by a
factorised distribution, replacing the full covariance matrix $\mA$ with a
vector $\va$ that contains only the variance of the $j=1,\ldots,P$ elements of
$\vx$,
\begin{equation}
  \label{eq:f_in_mf}
  \tilde{W}(\vx; \tilde{\va}, \tilde{\vb}) =  \prod_j^P \frac{1}{\tilde{Z}(\tilde{\eva}_j, \tilde{\evb}_j)}  p_X(\evx_j)
  \exp \left( \tilde{\evb}_j \evx_j -\frac{1}{2} \tilde{a}_{j} \evx_j^2 \right).
\end{equation}
where we use the tilde to denote mean-field quantities. We have implemented mean-field approximations for the models discussed thus far
and we show the performance of AMP with this approximation for the three models
discussed so far in \cref{fig:amp-vs-se_mf} together with the state
evolution prediction for the reconstruction of a single stored pattern $P=1$ without the
mean-field approximation. The picture that emerges is similar for all three
models studied here: the algorithm with the mean-field approximation is able to
reconstruct the stored patterns just as well as if it was looking at the reconstruction
of a single pattern up to a certain noise level, beyond which performance
quickly deteriorates. Intuitively, the cross-talk between the stored patterns introduces
an additional source of noise for the reconstruction, which leads to failure to
reconstruct the stored patterns at lower $\Delta$ than in the case $P=1$.

\paragraph*{Scaling of the critical number of stored patterns} Throughout this
paper, we have relied on the assumption that the matrix $\mJ$ is approximately
low-rank, in the sense that its eigenvalues can be separated into large bulk of eigenvalues, from a which only $P\sim\mathcal{O}(1)$ eigenvalues corresponding to the stored patterns detach. The mean-field approximation that we just
introduced makes it now computationally feasible to run the reconstruction
algorithm even with a large number of patterns. This raises an important
practical question: for which number of patterns does the algorithm based on the
low-rank approximation break down?

\begin{figure}[t]
  \centering
  \includegraphics[width=.48\linewidth]{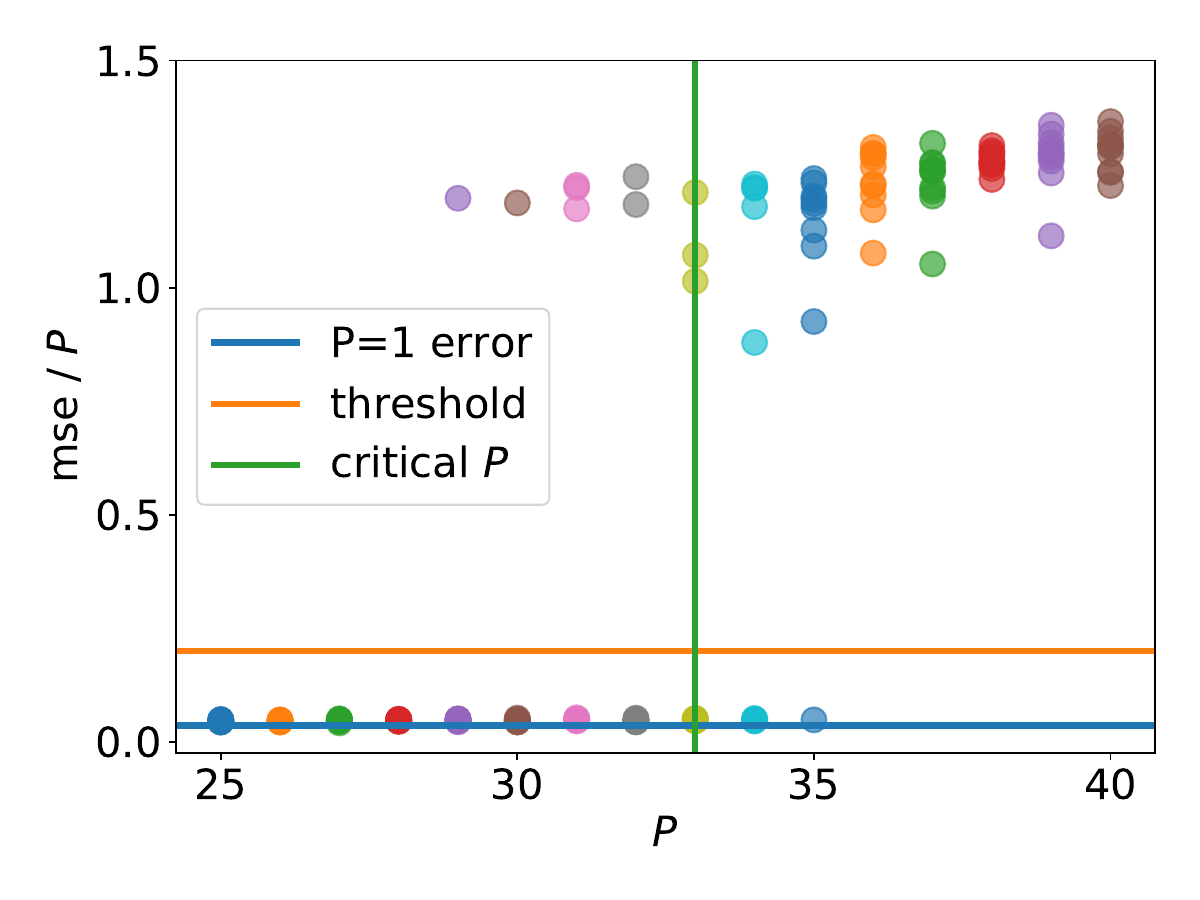}%
  \includegraphics[width=.48\linewidth]{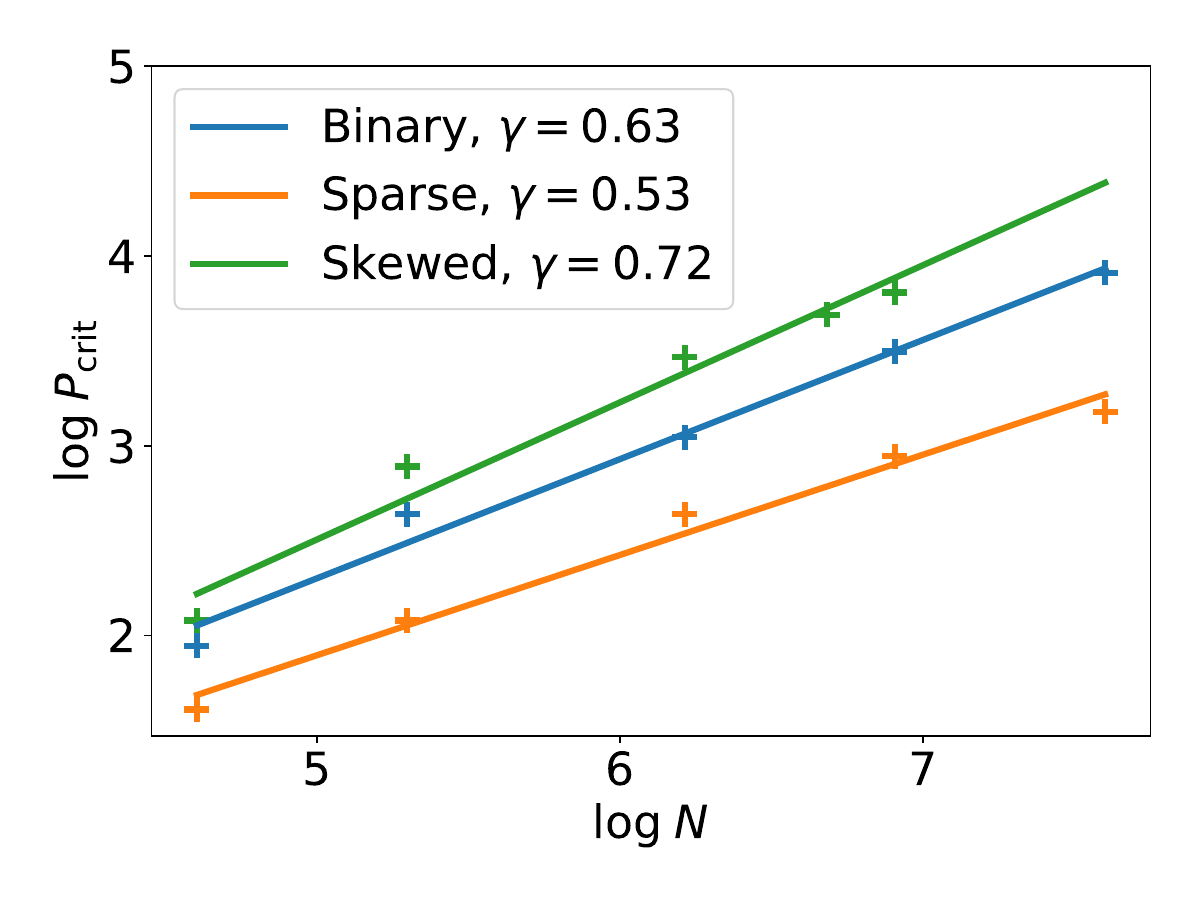}
  \caption{\label{fig:meanfield-scaling} \textbf{Scaling of the critical number
      of patterns that can reliably be reconstructed with mean-field message
      passing.} \figleft{} For binary patterns, we plot the final reconstruction
    error of twenty instances of the mean-field message-passing algorithm as a
    function of the number of stimuli stored $P$.  We define the critical number
    of patterns that can be reconstructed as the largest $P$ at which more than
    half the runs yield an $\mse$ better than a threshold value (here, $20\%$
    of the trivial $\mse$, i.e.~0.2). \emph{Parameters:}
    $N=1000, \Delta=0.2\Delta_C, \tau=0$. \figright{} We plot the highest number
    of patterns~$P_{\max}$ that could be reconstructed with an error below
    $20\%$ of the trivial error in at least~$50\%$ of cases. For all three
    models considered, we show experimental results (crosses) and the exponent
    of a power law fit $P_{\max}\sim N^\gamma$. In all cases, $\tau=0$ and the
    noise level $\Delta$ was $20\%$ of the critical noise level $\Delta_c$ where
    reconstruction becomes impossible. $\rho=0.3$ for both the sparse Hopfield
    and Tsodyk's prior.}
\end{figure}

We investigate this question in all three models numerically as follows. We
fixed the noise level $\Delta$ in all experiments at $20\%$ of the critical
noise level beyond which it is information-theoretically impossible to weakly
reconstruct the stored patterns. We then ran
twenty reconstruction experiments with~$N=1000$ for each value of $P$; the final $\mse$ for each run is shown with a dot on the left of \cref{fig:meanfield-scaling}. For
$P=25$, all twenty runs gave an $\mse$ well below the threshold; in fact, the
reconstruction error per pattern $\mse / P$ is as low as if we had stored only a
single pattern in the network (blue line). As we increase the number of patterns
$P$, the first time the algorithm is not able to recover all patterns is for
$P=29$ patterns. For $P=36$ patterns, the algorithm did not reconstruct the patterns with an error better than chance a single time. Given the clear separation between successful runs with low error and unsuccessful runs with an error that is essentially random guessing, we set the threshold for the reconstruction $\mse$
below which we consider the algorithm successful at~$20\%$ of the trivial $\mse$
obtained by random guessing (orange line).  At $P=34$, the algorithm fails to achieve an $\mse$ below the
threshold in more than $50\%$ of the cases. We define the critical number of
stored patterns $P_{\text{crit}}$ as the largest number of patterns that can be
reconstructed with an $\mse$ below the threshold in at least $50\%$ of the runs,
so $P_{\text{crit}}=33$ in this case. On the right, we show the values
of~$P_{\text{crit}}$ for all three models (binary, sparse and skewed) as a
function of $N$. In each case, we find that $P_{\text{crit}}\sim N^\gamma$, with
the exponent $\gamma$ between $\gamma \approx 0.5$ for sparse patterns
and~$\gamma \approx 0.7$ for patterns from Tsodyk's prior. The exact values of the exponents will depend on the choices of the threshold and especially the noise level. While the separation between successful and unsuccessful runs allows for various error thresholds without changing the results, the exponents are more sensitive to the noise level, since higher noise levels induce higher fluctuations in the reconstruction errors closer to the critical noise (cf. \cref{fig:amp-vs-se_mf}).

\section*{Discussion}
\addcontentsline{toc}{section}{Discussion}

We have derived the conditions under which it is possible to reconstruct the
patterns stored in a recurrent Hopfield-type network from knowledge of the
connections alone, in a case where those weights are obtained by a learning rule
which is a noisy, non-linear version of the classic Hebbian rule. We have
implemented and provided practical algorithms to do so: a Bayesian approach
based on message passing, and classic non-parametric approaches such as PCA on
the weights and on the Fisher score matrix. The message-passing algorithm offer
a principled way to reconstruct the patterns using prior information about the
prior of the inputs, while the spectral methods are robust and
easy-to-implement, but fail to take this extra information into account.

The performance of the algorithms can be captured by an effective noise level,
which takes both the synaptic noise and the thresholding of the learning rule
into account. We found that the message-passing algorithm
beats the PCA reconstruction across noise levels, which is due to the principled
way in which the message-passing algorithm can incorporated prior knowledge
about the distribution of patterns.

While our theoretical results were obtained in the limit where the number of
patterns stored is of order 1, while the number of neurons in the network tends
to infinity, we have also explored a mean-field approximation that allowed us to
study how the maximum number of patterns that can be reconstructed with some
reliability scales with the pattern dimension. We now discuss some directions in
which our work can be extended.

\paragraph*{The case of unknown generative model} We have assumed that the
hyper-parameters of the problem, such as the number of stored stimuli $P$ or the
generative model of the stimuli $p_X(\vx)$, are known before
reconstruction. However, it is possible to extend both the message-passing
algorithm and its analysis to the case of unknown hyper-parameters. Within our
framework, a natural approach to learn the values of these hyper-parameters is
closely related to the well-known expectation maximisation
algorithm~\cite{dempster1977maximum}. A detailed study of expectation
maximisation to obtain various hyper-parameters for message-passing algorithms
was reported by Krzakala et al.~\cite{krzakala2012probabilistic} for the
statistical estimation problem of ``compressed sensing'', where one aims to
reconstruct a signal from a number of measurements that is smaller than the
number of unknowns. Another approach to deal with an unknown noise distribution
was recently proposed by Montanari et al.~\cite{montanari2018adapting} for the
related problem of matrix denoising, where one aims to reconstruct an unknown,
low-rank matrix $\mX \in \mathbb{R}^{m\times n}$ from observations
$\mY = \mX + \mW$ when $\mW$ is a noise matrix with independent and identically
distributed entries. Montanari et al.~\cite{montanari2018adapting} propose an
iterative approach to estimate the noise distribution from the observations
$\mY$. An interesting direction for future work is to explore both strategies
for the problem of reconstructing stored patterns in connectivity matrices.

From the point of view of the theoretical analysis, assuming the ``wrong'' rank in the algorithm in the sense that there is a mismatch between the number of patterns one wants to infer and the number of patterns stored makes the analysis much more involved. The message-passing algorithms can be modified to take this mismatch into account. While first steps towards a theoretical analysis and the development of modified message-passing have recently been made~\cite{antenucci2019glassy, antenucci2019approximate, lucibello2019generalized}, we leave the exploration of this direction in the context of neuroscience to future work.

Finally, there could also be no (discernible) signal in the connectivity matrix, because the relative strength of the noise is too large, or because the connectivity is purely random. As we discussed above, AMP shows a clear advantage over PCA in this case: while AMP will return inputs which are close to 0, indicating that it didn't find any structure in the data, PCA will always return the leading eigenvector, which in this case however is completely uninformative.

\paragraph*{Reconstructing an extensive number of patterns} Another important
extension of our work would be the reconstruction of an extensive number of
patterns from the connectivity $\mJ$, \emph{i.e.} whether the regime
$P\sim \mathcal{O} (N)$ is accessible. This is essentially the problem of
factorising a large matrix with extensive rank, which is also known as
dictionary learning~\cite{mallat1993matching, elad2006image, mairal2010online}.
While we note that there have been promising signs of progress
recently~\cite{barbier2021statistical, maillard2021perturbative}, this remains a
hard problem with implications that would go far beyond the application
discussed here.

\paragraph*{Reconstructing the patterns from the dynamics} It may be possible to retrieve a pattern given a distorted or noisy version of it by initialising the
network with this distorted pattern and running the network dynamics~\eqref{eq:dynamics} until
convergence. Another algorithm to reconstruct the patterns is then to run the network dynamics from different initial conditions until convergence, and to take the resulting fixed points as estimates of the patterns stored in the connectivity. 
However, this procedure will only yield stored patterns if the
initial condition is correlated with one of the patterns stored in the
network. It is thus more apt to speak about ``pattern completion'' in this
context~\cite{mackay2003}. Here, we considered the more challenging problem where
such partial observations of the patterns are not available, which would be the
case if the only information available is the connectivity matrix. 

It is still fair to wonder how well this algorithm could do when starting from random initial conditions. In the classical Hopfield model, where the weights are given by the Hebb rule~\eqref{eq:w}, any initial condition of the dynamics that has non-zero macroscopic overlap with a stored pattern is guaranteed to converge to that pattern when the number of patterns is of order one. In a finite network, one then expects random initial conditions to converge to one of the stored patterns (depending on the random initial overlaps). While we study a learning rule that is different from Hebb's rule, we note a classic result by
Sompolinsky~\cite{sompolinsky1986neural} that suggests that the dynamics of recurrent
networks whose connections are a nonlinear, noisy version of the Hebbian
weights~\eqref{eq:w} are largely unaffected by this change. 

For a single pattern, iterating the dynamics resembles the power method~\cite[sec.~8.5]{horn2013matrix} to compute the first principal component of a symmetric matrix, and the dynamical approach to reconstruction thus closely resembles the PCA algorithm that we analysed in the case $P=1$.  For more than one pattern, there is no guarantee however that \emph{all} the patterns will be recovered. Moreover, the performance of this algorithm is expected to be strongly model-dependent: in more realistic models such as the Tsodyks-Feigel'man model~\cite{tsodyks1988a, tsodyks1988b}, the quiescent state, where all neurons are at zero, is stable, and convergence to one of the stored patterns will necessitate an initial overlap that is larger than some non-zero critical value. Recovering the patterns through the dynamics is not expected to work in this case.

Similarly, this strategy is also expected to fail when reconstructing an extensive number of patterns. In this case, we need a correlation between the initial condition and each pattern which is bounded away from zero, and obtaining such a correlation for all the patterns would require exponentially many trials.

\paragraph*{What information on synaptic weights does connectomic data provide?} EM reconstructions of neuronal circuits are of sufficiently high resolution that they can enable measurements of the volume of dendritic spines, the anatomical structures on which the vast majority of synaptic connections between pyramidal cells are formed (see e.g.~\cite{dorkenwald19}). The volume of dendritic spines has in turn be shown to be strongly correlated with synaptic strength, as measured by the amplitude of post-synaptic potentials triggered by pre-synaptic activity \cite{matsuzaki01, noguchi11, holler2021structure}. Noise in measurement of dendritic spines, and the lack of perfect correlation between volume and EPSP amplitudes, are two of the reasons (together with noise in the learning process) for the introduction of the noise matrix $\zeta_{ij}$ in our model.

\paragraph*{Towards more biologically realistic models} Our focus here was to
analyse the simplest model where the problem of retrieving the patterns is
mathematically well-posed, and neither trivial nor impossible to achieve. We now discuss various extensions of this model that could be addressed in future work.

{\it Asymmetry in the learned component of the connectivity matrix.} We have focused here on a symmetric connectivity matrix, for multiple reasons. First, multiple in vitro studies in both cortex and hippocampus have shown that local networks in these structures exhibit a significant degree of symmetry, as evidenced by a much higher probability of bidirectional connections in pairs of neurons, compared to a random directed Erdos-Renyi network \cite{markram97,song05,wang06,guzman16}, with the notable exception of rodent barrel cortex where no such overrepresentation exists \cite{lefort09}. In addition, synaptic plasticity in area CA3 of the hippocampus has been shown to be temporally symmetric \cite{mishra16}. One thus expects such synaptic plasticity rules to lead to connectivity matrices with a strong degree of symmetry in this area. In cortex, plasticity rules are temporally asymmetric as a function of the timing difference between spikes of pre and post-synaptic neurons. However, plasticity depends also strongly on firing rates of pre and post-synaptic neurons, and if dependence on firing rate dominates over the dependence on spike timing as has been suggested \cite{graupner16}, then one also expects a strongly symmetric Hebbian component in connectivity matrices. Finally, a strong degree of symmetry is also consistent with the observation of persistent activity in multiple cortical areas in rodents \cite{inagaki19} and primates \cite{fuster71,miyashita88,funahashi89,romo99}. Consistent with this idea, Inagaki et al \cite{inagaki19} showed using optogenetic perturbations that the dynamics in area ALM of the mouse during a short-term memory task exhibit multiple characteristics of attractor dynamics. Of course, we do not expect connectivity matrices in all brain structures to be well captured by noisy symmetric matrix. In particular, networks storing temporal sequences \cite{kleinfeld88, fiete10, gillett20} would need to contain a significant asymmetric learned component. Our methods would then need to be extended to asymmetric matrices. From the practical point of view, the application of the method proposed makes only sense when the reconstructed connectivity matrices exhibit a significant degree of symmetry.

{\it Inhibition}. Here, we have assumed inhibition is not involved in learning and simply provides a uniform inhibition to excitatory neurons, equivalent to setting a threshold for active neurons. This traditional view is consistent with the observation of high connection probabilities between a specific type of inhibitory neuron (PV positive interneurons) and pyramidal cells \cite{fino11}. It is also consistent with the observation in hippocampus that inhibition is only weakly modulated by spatial location \cite{grienberger17} (see however \cite{rolotti22}). However, it has also been shown that synaptic connections involving inhibitory interneurons also exhibit plasticity \cite{froemke15}, and it has been argued that plasticity of such connections could greatly expand storage capacity \cite{mongillo2018inhibitory}. Our methods could be extended to the addition of inhibitory neurons, with a few caveats: estimate of the strength of such synapses might be more challenging, since synapses involving interneurons are formed directly on dendritic shafts and not on spines; Also, the connectivity matrix will then necessarily be asymmetric. 

{\it Memories stored with different strengths}. In our model, as in most associative memory models, memories are stored with identical strengths. A straightforward extension of the model would be a model where memories have different embedding strengths. This class of models include `palimpsest' models in which recently stored patterns gradually erase older patterns that are progressively forgotten \cite{mezard86, parisi86, amit94b, lahiri13}. Note that in this class of models, our method would be likely to infer only the most strongly embedded patterns, and memories that are on their way to becoming forgotten would not be likely to be inferred. From the point of view of message-passing algorithms, the gradual erasure would have to be modelled through the learning rule of the model, Eq.~\eqref{eq:J}, together with a prior over embedding strengths. 

{\it Distributions of synaptic weights}. Our model leads to a truncated Gaussian
distribution of non-negative synaptic weights. Distributions of experimentally
recorded EPSP amplitudes \cite{song05} as well as spine volumes~\cite{loewenstein11} have been fitted using log-normal distributions. Our method
can easily be generalized to networks with arbitrary distributions of
non-negative weights, by using a suitable non-linearity $\Phi$ in \cref{eq:J}~\cite{mongillo2018inhibitory}. 

{\it Binary Hebbian matrices}. It has been proposed by some authors that synapses store information in a digital, not analogue fashion \cite{sompolinsky1986neural,petersen98,oconnor05}. In this scenario, synapses have only a few stable states, and plasticity events correspond to transition between these states. The resulting model would then bear similarities with stochastic block models (SBMs)~\cite{holland1983stochastic}, where groups of neurons representing a particular stored memory would correspond to communities in SBMs. An important difference is that in the case of random patterns, there would be overlaps between these groups~\cite{airoldi2008mixed}. One could use similar methods as the ones proposed here to deal with such a scenario, since recovering communities in stochastic block models can be reformulated as a low-rank matrix factorisation problem, and is hence amenable to the same analysis techniques that we used here, see refs.~\cite{decelle2011inference, Lesieur2017a, bandeira2018notes} for examples of this rich literature.

{\it Distributions of patterns}. We have assumed that stored memories are random i.i.d. binary vectors. Responses of neurons to external inputs rarely follow a bimodal distributions, and can sometimes be better described by a unimodal lognormal distribution \cite{lim15}. However, non-linearities associated with the synaptic plasticity rule could potentially binarize stored memories \cite{pereira2018attractor}, which would then result in a model that is very similar to the model investigated here.


\paragraph*{Validation.} An important question is how to validate the results of such an analysis.
One possibility would be simulate the dynamics of a network whose connectivity matrix is set to be the reconstructed matrix, using the inferred patterns as initial conditions, to check that the dynamics converges to fixed point attractors that are close to the inferred patterns.

\section*{Methods}
\addcontentsline{toc}{section}{Methods}

\subsection*{A formal analogy between inference problems and statistical physics}
\addcontentsline{toc}{subsection}{A formal analogy between inference problems and statistical physics}

It may come as a surprise that statistical physics can be helpful in solving and
analysing inference problems like the one considered here. The connection
between the two becomes more transparent if we introduce the interaction
$g(\cdot) \equiv \ln P_{\out}(\cdot)$ to rewrite the posterior as
\begin{equation}
  \label{eq:gibbs}
  P(\mX|\mJ) = \frac{1}{Z(\mJ)} \prod_i^N P_X(\vx_i) \prod_{j>i}^N \exp\left[ g(\emJ_{ij}, \emW_{ij}) \right].
\end{equation}
This distribution describes the posterior density over estimates $\mX$. However,
it can also be interpreted as the Gibbs or Boltzmann distribution that describes
the properties of complex, disordered systems such as glasses. This analogy can
be leveraged by exploiting tools from the statistical physics of disordered
systems to tackle the -- hard -- inference problem that is inferring the
patterns from the connectivity $\mJ$. The key ideas behind the AMP
algorithm that we discuss throughout this work and AMP algorithms in general
first appeared in a paper by Thouless, Anderson and Palmer~\cite{Thouless1977}
that dealt with physical systems described by an equilibrium distribution of the
type~\eqref{eq:gibbs}. State evolution techniques where first introduced for
compressed sensing problems by Donoho, Maleki and Montanari~\cite{Donoho2009}
based on ideas from~\cite{Bolthausen2014}, but it too is based on ideas from
statistical physics often referred to as the cavity
method~\cite{Mezard1986}. For a much more detailed on the links between
statistical physics and inference problem,
see~\cite{mackay2003,Mezard2009,zdeborova2016}.

\subsection*{Approximate message passing for low-rank matrix reconstruction}
\addcontentsline{toc}{subsection}{Approximate message passing for low-rank matrix reconstruction}

We are now in a position to state the AMP algorithm for the factorisation
of symmetric low-rank matrices that in this form was derived by
Lesieur~\cite{Lesieur2017a}, building on the previous works deriving AMP-type
algorithms for particular instances of this problem
class~\cite{fletcher2012,matsushita2013a,Deshpande2014}. We refer the interested
reader to these papers for details on the derivation of this class of algorithms
and their relation to belief propagation and spectral methods.

To describe the algorithm, we first define the Fisher score matrix as a
transformation of the data matrix $\mJ$ given for the inference:
\begin{equation}
  \emS_{ij}(\emJ_{ij}) \equiv \left.\frac{\partial \ln P_{\mathrm{out}}(\emJ_{ij} |
      w)}{\partial w} \right\rvert_{w=0}
\end{equation}
For the channel corresponding to the rectified Hopfield
model~\eqref{eq:J}, we find 
\begin{equation}
  \label{eq:S}
  \emS_{ij}(J_{ij})  = \begin{cases} - 
    \dfrac{2e^{-\nicefrac{\tau^2}{2\nu^2}}}{\sqrt{2\pi}\,\nu\erfc\left(\nicefrac{-\tau}{\sqrt{2}\nu}\right)},
    & \emJ_{ij} = 0 \\[1.5em]
    \dfrac{J_{ij} + \tau}{\nu^2} & \text{otherwise}.
  \end{cases}
\end{equation}

Low-RAMP is an iterative algorithm: at every step $t$, it computes a new
estimate of the mean $\hat{\vx}_i^{t+1}$ and the variance $\sigma_i^{t+1}$ as 
\begin{subequations}
  \label{eq:amp_tap}
  \begin{align}
    \hat{\vx}_i^{t+1} & = f(\mA^t_i, \vb^t_i) \label{eq:amp_tap_x}\\
    \sigma_i^{t+1} & = \partial_\vb f(\mA^t_i, \vb^t_i)\label{eq:amp_tap_sigma}
  \end{align}
\end{subequations}
where the threshold function was defined in \cref{eq:fin} and is repeated
here for convenience:
\begin{equation}
  f(\mA, \vb) =  \frac{1}{Z(\mA, \vb)} \sum_{\vx\in\mathcal{X}^P} \vx\, p_X(\vx) \exp \left( \vb \vx -
    \frac{1}{2}\vx^\top \mA \vx \right).
\end{equation}
There exists a set of parameters $\mA_i\in\mathbb{R}^{P\times P}$ and
$\vb_i\in\mathbb{R}^P$ for every marginal, which in turn are updated as
\begin{subequations}
  \begin{align}
        \vb_i^t & = \frac{1}{\sqrt{N}} \sum_k^N \emS_{ki} \hat{x}^t_k -
            \left(\frac{1}{N} \sum_k^N S_{ki}^2 \sigma_k^t \right)\hat{x}^{t-1}_i \label{eq:amp_tap_B}\\
    \mA_i^t & = \frac{1}{N} \sum_k^N S_{ki}^2 \hat{x}^t_k
            \hat{x}^{t,\top}_k \label{eq:amp_tap_A}
  \end{align}
\end{subequations}  

To run the algorithm, we perform these steps:
\begin{enumerate}
\item Given the matrix $\mJ$, compute the Fisher score matrix $\mS$ using
  \cref{eq:S}.
\item For all $i=1,\ldots,N$, initialise the parameters $\vb_i$ and $\mA_i$ such
  that all entries are zero. Initialise all estimators $\hat{\vx}^t_i$ with a
  random draw from the prior distribution $p_X(\vx)$ and set $\hat{\vx}^{t-1}_i$
  to all zeros for the first step. (There is no need to initialise $\sigma_i$).
\item Compute first the update to $\mA_i^t$ and $\vb_i^t$ following
  Eqs.~\eqref{eq:amp_tap_A} and~\eqref{eq:amp_tap_B}, then compute the new
  means~$\hat{\vx}^{t+1}_i$ and their variance~$\sigma_i^{t+1}$ using
  Eqs.~\eqref{eq:amp_tap_x} and~\eqref{eq:amp_tap_sigma}.
\item Repeat Step~3 until the squared difference between all $\hat{x}^t_i$ and
  $\hat{x}^{t+1}_i$ is smaller then some predefined threshold $\epsilon$.
\end{enumerate}
We also provide an implementation of this algorithm in a
\href{https://github.com/sgoldt/reconstructing_memories}{Python package} that
was the base of all the programs written for this paper.

\subsection*{State evolution for reconstructing sparse patterns}
\addcontentsline{toc}{subsection}{State evolution for reconstructing sparse patterns}

We discussed in the main text that the sparse prior~\eqref{eq:prior_hf-sparse}
has mean $\langle x \rangle=0_p$ and covariance
$\langle x x^\top \rangle=\rho \mI_p$. The state evolution will interpolate
between an order parameter that is all zeros at initialisation, and
$M=\rho \mI_P$ for perfect reconstruction. We are thus motivated to use the
ansatz $M^t = m^t \mI_P$. The threshold function then becomes
\begin{equation}
  [{f(A, B)}_k] = \frac{\rho e^{-A_{kk}/2}\sinh(B_k)}{1 +
    \rho\left[e^{-A_{kk}/2}\cosh(B_k) -1\right]}
\end{equation}
Substituting this form into the SE update equation~\eqref{eq:se} yields a closed
update equation for the parameter $m^t$,
\begin{equation}
  \label{eq:rhf_se}
  m^{t+1} = \underset{w}{\mathbb{E}} \; \frac{\rho^2 e^{-m^t/2\Delta}
    \sinh\left( m^t/\Delta + \sqrt{m^t/\Delta} z \right)}{1
    + \rho\left[ e^{-m^t/2 \Delta} \cosh\left(m^t/\Delta + \sqrt{m^t/\Delta} z\right) -1\right]},
\end{equation}
where $z$ is again a scalar Gaussian random variable with zero mean and unity
variance, like in \cref{eq:se-P1}. We can recover the state evolution for the
symmetric rectified Hopfield model from \cref{eq:rhf_se}, in the limit
$\rho\to1$.

\subsection*{State evolution for reconstructing patterns with low coding level}
\addcontentsline{toc}{subsection}{State evolution for reconstructing patterns with low coding level}

For Tsodyk's prior~\eqref{eq:prior_tsodyks}, we have a new threshold
function~\eqref{eq:fin} which reads
\begin{equation}
  \label{eq:fin_tsodyks}
  [{f(A, B)}_k] = 1 - \rho-\frac{1-\rho }{1-\rho  \left(1-e^{A_{kk} \left(\rho
          -\nicefrac{1}{2}\right)+B_k}\right)}
\end{equation}
The prior distribution~\eqref{eq:prior_tsodyks} allows us to use the same ansatz
for the magnetisation that we used above to analyse sparse patterns. Setting
$M^t = m^t \mI_P$, we get the following update equation for the scalar order
parameter $m^t > 0$:
\begin{equation}
  \label{eq:hf_tsodyks_se}
  m^{t+1} = \underset{W}{\mathbb{E}} \; \frac{{(\rho -1)}^2 \rho ^2 \left(e^{m^t/\Delta}-1\right) e^{w
      \sqrt{m^t/\Delta}}}{\left(\rho  e^{w \sqrt{m^t/\Delta }}-(\rho -1)
      e^{\nicefrac{m^t}{2 \Delta }}\right) \left(\rho \exp\left( \frac{2 w \sqrt{\Delta 
            m^t}+m^t}{2 \Delta } \right)-\rho+1\right)}.
\end{equation}

\section*{Acknowledgements}
\addcontentsline{toc}{section}{Acknowledgements}

SG, FK and LZ would like to thank the Department of Mathematics at Duke
University for their hospitality during an extended visit.




\printbibliography

\end{document}